\documentclass{autart}
\usepackage[dvipsnames]{xcolor}

\usepackage{graphicx}
\usepackage{amsmath}
\usepackage{amssymb}

\usepackage{amsfonts}
\usepackage{dsfont} 
\usepackage{subfigure}
\usepackage{algorithm}
\usepackage[noend]{algpseudocode}
\usepackage{nicefrac}
\usepackage{mathrsfs}
\usepackage{enumitem}
\usepackage{natbib}

\renewcommand{\qed}{\hfill$\blacksquare$}
\newcommand{\qedwhite}{\hfill \ensuremath{\Box}}

\usepackage{xcolor}
\definecolor{forestgreen}{rgb}{0.13, 0.55, 0.13}
\definecolor{orange}{rgb}{1,0.49,0}

\newtheorem{defn_new}{Definition}[section]
\newtheorem{thm_new}{Theorem}[section]
\newtheorem{lem_new}{Lemma}[section]

\newtheorem{cor_new}{Corollary}[section]

\newtheorem{rem_new}{Remark}[section]
\newtheorem{exam}{Example}[section]
\newtheorem{assump}{Assumption}[section]
\newtheorem{prob_new}{Problem}[section]

\newcommand{\R}{\mathbb{R}}
\newcommand{\N}{\mathbb{N}}
\newcommand{\C}{\mathcal{C}}
\newcommand{\X}{\mathcal{X}}

\newcommand{\G}{\mathcal{G}}
\newcommand{\V}{\mathcal{V}}
\newcommand{\W}{\mathcal{W}}
\newcommand{\E}{\mathcal{E}}
\newcommand{\D}{\mathcal{D}}
\newcommand{\Sig}{\mathcal{S}}
\newcommand{\OO}{\Omega}

\newcommand{\sat}{\vDash}
\newcommand{\F}{\mathcal{F}}
\newcommand{\HH}{\mathcal{H}}

\newcommand{\BR}{\mathrm{BR}}

\begin{document}
\begin{frontmatter}

\title{Contract Composition for Dynamical Control Systems:\\ Definition and Verification using Linear Programming}


\author[KTH]{Miel~Sharf},~
\author[RUG]{Bart~Besselink},~
\author[KTH]{Karl~Henrik~Johansson}

\address[KTH]{Division of Decision and Control Systems, KTH Royal Institute of Technology, and Digital Futures. 10044 Stockholm, Sweden (e-mail: {\rm\{sharf,kallej\}@kth.se)}.}
\address[RUG]{Bernoulli Institute for Mathematics, Computer Science and Artificial Intelligence, University of Groningen, 9700 AK Groningen, The Netherlands (e-mail: {\rm b.besselink@rug.nl}).}

\begin{abstract}                
Designing large-scale control systems to satisfy complex specifications is hard in practice, as most formal methods are limited to systems of modest size. Contract theory has been proposed as a modular alternative to formal methods in control, in which specifications are defined by assumptions on the input to a component and guarantees on its output. However, current contract-based methods for control systems either prescribe guarantees on the state of the system, going against the spirit of contract theory, or can only support rudimentary compositions. 
In this paper, we present a contract-based modular framework for discrete-time dynamical control systems. We extend the definition of contracts by allowing the assumption on the input at a time $k$ to depend on outputs up to time $k-1$, which is essential when considering the feedback connection of an unregulated dynamical system and a controller. 
We also define contract composition for arbitrary interconnection topologies, under the pretence of well-posedness, and prove that this notion supports modular design, analysis and verification. This is done using graph theory methods, and specifically using the notions of topological ordering and backward-reachable nodes. Lastly, we use $k$-induction to present an algorithm for verifying vertical contracts, which are claims of the form ``the conjugation of given component-level contracts is a stronger specification than a given contract on the integrated system". These algorithms are based on linear programming, and scale linearly with the number of components in the interconnected network. A numerical example is provided to demonstrate the scalability of the presented approach, as well as the modularity achieved by using it.
\end{abstract}

\begin{keyword}
Formal methods, contracts, linear programming, modular design, graph theory, interconnection topology
\end{keyword}

\end{frontmatter}

\section{Introduction}\label{sec.Intro}
\vspace{-17pt}
In recent years, modern engineering systems have become larger and more complex than ever, as the number of different components and subsystems is rapidly increasing due to the prominence of the ``system-of-systems" design philosophy. At the same time, these systems are subject to specifications with constantly increasing intricacy, including safety and performance specifications. As a result, the validation and verification process, which must be conducted before deployment, has become exponentially more difficult. Recently, several attempts have been made to adapt contract theory, which is a modular approach for software design, to dynamical control systems. In this paper, we present a modular approach for contract-based design of dynamical control systems by defining a ``contract algebra", considering the composition of contracts on different components with a general interconnection topology. We prove that our definition supports independent design, analysis, and verification of the components or subsystems. We also prescribe linear-programming (LP)-based tools for verifying that a given contract on the integrated system is implied by a collection of component-level contracts.

\vspace{-8pt}
\subsection{Background}
\vspace{-10pt}
Modularity is a widely accepted philosophy of system design. Identifying a natural partition of a large-scale system into smaller modules enables independent and parallel work on the different components by different teams, as well as outsourcing part of the work to a subcontractor. Modular design also supports future modifications in the design, as only the updated components need to be re-verified rather than the entire system. For these reasons, a wide range of literature advocates for designing large-scale systems using as much modularity as possible, see \cite{Baldwin2006} and \cite{Huang1998} for discussions on modular design in engineering systems and electromechanical consumer products. The opposite approach, known as integral design, in which a single designer integrates all parts of the system, should also be recalled [\cite{Ulrich1995}].

Modular design and verification techniques are lacking for dynamical control systems. Safety is most commonly defined via controlled invariant sets [\cite{Blanchini2008}], but can only handle rudimentary safety specifications, and cannot be applied modularly.
Existing modular techniques, such as dissipativity theory [\cite{Willems1972a,Willems1972b}], can only handle limited performance specifications, and cannot be used for safety.
In contrast, formal methods in control, which are adaptations of automata-based model-checking algorithms in software engineering [\cite{Baier2008}], provide verification methods and correct-by-design synthesis procedures for specifications given by temporal logic formulae [\cite{Belta2007,Tabuada2009}]. However, they are integral design methods which scale exponentially with the dimension of the system, and are thus applicable only to systems of modest size. Also, most works on scalable distributed and decentralised control methods, such as \cite{Siljak2005} and \cite{Rantzer2015}, are not modular, as they require a single authority with complete knowledge of the system model to design the decentralised or distributed controllers.

Lately, several modular approaches have been proposed to tackle problems in the design of dynamical control systems. One example is composition-compatible notions of abstraction and simulation, attempting to ``modularise" formal methods in control [\cite{Zamani2018, Saoud2018b}]. Another approach attempts to relate controlled-invariant sets and reachability analysis on the subsystem-level to controlled-invariant sets and reachability analysis on the composite system-level [\cite{Smith2016,Chen2018}]. A third approach, and the focus of this paper, is contract theory.
Contract theory is the most prominent software-theoretic modular design philosophy [\cite{Meyer1992,Benveniste2018}]. It explicitly defines assumptions on the input and guarantees on the output of each software component, providing methods for design and verification of software packages. Contract theory hinges on the notions of satisfaction, refinement and composition, allowing implementation, comparison and conjugation of contracts, as well as computationally-viable tools for verifying these notions. 

Several recent attempts have been made to apply contract theory in the realm of dynamical control systems. The works of Nuzzo et al. apply contract theory to the ``cyber" aspects of cyber-physical systems, see \cite{Nuzzo2015,Nuzzo2014} and references therein. More recently, other attempts have been made to apply it to dynamical control systems. The papers \cite{Besselink2019} and \cite{Shali2021} focus on continuous-time systems, and use verification methods based on geometric control theory and behavioural systems theory, respectively. In contrast, the works \cite{Saoud2018,Eqtami2019} and \cite{Ghasemi2020} focus on discrete-time systems, prescribing contracts with assumptions on the input signal to the system, and guarantees on the state and the output of the system. The latter is a limitation in contract theory, as the state of the system is an internal variable that should not be a part of its interface. This is also the case for \cite{Saoud2021}, which considers continuous-time systems. This problem was remedied in \cite{SharfADHS2020} and \cite{Sharf2021b}, which consider contracts with guarantees on the output relative to the input, supporting the incorporation of sensors and other systems in which the guarantees on the output should depend on the input. The paper \cite{SharfADHS2020} presented preliminary LP-based tools for verifying satisfaction, which were significantly extended in \cite{Sharf2021b}. However, only a rudimentary notion of composition was considered, merely defining the cascade composition of two contracts, without providing any associated computational tools.

\vspace{-8pt}
\subsection{Contributions}
\vspace{-10pt}
This paper develops a modular and compositional framework based on contract theory for discrete-time control systems. These results extend considerably existing methods in the literature, allowing the assumption on the input at time $k$ to depend on the values of the output up to time $k-1$. These contracts arise naturally when considering feedback control, as the control input to the dynamical system should depend on the current output. We then define contract composition for arbitrary network interconnections, and provide LP-based algorithms for verifying that an interconnection of component-level contracts refines a contract on the integrated system. These are first achieved for networks without feedback loops (Definition \ref{defn.Feedbackless} and Algorithm \ref{alg.VerifyNSysCascade}), and are later generalised to arbitrary well-posed network interconnections (Definition \ref{def.FeedbackComp} and Algorithm \ref{alg.VerifyNSysFeedback}). In each case, we prove that the composition supports modular design, analysis, and verification. Moreover, we prove the presented algorithms are always correct, and that they scale linearly with the number of components in the integrated system. These contributions, together with the results of \cite{Sharf2021b}, provide the first true adaptation of contract theory for discrete-time dynamical control systems, providing a modular framework for satisfaction, refinement, and composition, all supported by tractable LP-based computational tools.

The paper is organised as follows. Section \ref{sec.Background} presents required background on contract theory and graph theory. Section \ref{sec.ProbForm} introduces generalised contracts, as well as a formal definition of the problems discussed in the paper. Section \ref{sec.Feedbackless} considers feedback-less networks, and Section \ref{sec.GeneralFeedback} considers general well-posed networks. Section \ref{sec.CaseStudy} applies these methods in a numerical example.

\vspace{-15pt}
\paragraph*{Notation}
Let $\N = \{0,1,\ldots\}$ be the set of natural numbers. For $n_1,n_2 \in \N$, we let $\mathcal{I}_{n_1,n_2} = \{n_1,\ldots,n_2\}$ if $n_1 \le n_2$, and $\mathcal{I}_{n_1,n_2} = \emptyset$ otherwise. The collection of discrete-time signals $\N \to \R^d$ will be denoted by $\Sig^d$. A coordinate-projection matrix $P\in \R^{n_{d_1}\times n_{d}}$ is a matrix achieved by choosing a subset of rows from the identity matrix ${\rm Id}_{n_{d}}$ such that $P$ maps any vector $v\in \R^{n_{d}}$ to a vector composed of a subset $\mathcal{J}_P \subseteq \mathcal{I}_{1,n_d}$ of its entries. Moreover, its complementary coordinate projection matrix corresponds to the subset $\mathcal{I}_{1,n_d}\setminus\mathcal{J}_P$. We say that $d_1 \in \Sig^{n_{d_1}}$ is a subsignal of $d \in \Sig^{n_d}$ if there exists a coordinate-projection matrix $P\in \R^{n_{d_1}\times n_d}$ such that $d_1(k) = Pd(k)$ for any time $k\in \N$. We say that $d_2$ is the complementary subsignal to $d_1$ if $d_2(k) = Qd(k)$, where $Q$ is the complementary coordinate projection matrix to $Q$. In other words, any coordinate of the signal $d$ either belongs to $d_1$ or to $d_2$.
For a signal $v \in \Sig^{m}$ and $k_1,k_2\in \N$, we denote the vector containing $v(k_1),v(k_1+1),\ldots,v(k_2)$ as $v(k_1:k_2) \in \R^{(k_2-k_1+1)m}$. A set-valued map $f: X\rightrightarrows Y$ between two sets $X,Y$ associates a subset $f(x) \subseteq Y$ to any element $x\in X$. Moreover, $X^n$ is the set of $n$-tuples of elements of $X$. For vectors $u,v\in \R^n$, we write $u\le v$ if and only if $u_i \le v_i$ holds for any coordinate $i\in \mathcal{I}_{1,n}$.
%
The variables $k,\ell,n$ denote times in $\N$, and $p,q$ denote numbers in the set $\mathcal{I}_{1,N}$.
\vspace{-10pt}

\section{Background} \label{sec.Background}
\vspace{-7pt}
This section presents required background material on contract theory and graph theory.
\vspace{-7pt}
\subsection{Systems and Assume/Guarantee Contracts}
\vspace{-7pt}
We first define the class of systems we consider, which are seen as operators on the set of all possible signals.
\vspace{-7pt}
\begin{defn_new} \label{def.IOMap}
A (dynamical) system $\Pi$ with input $d\in \Sig^{n_d}$ and output $y\in\Sig^{n_y}$ is a set-valued map $\Pi : \Sig^{n_d} \rightrightarrows \Sig^{n_y}$. In other words, for any input trajectory $d\in \Sig^{n_d}$, $\Pi(d)$ is the set of all corresponding output trajectories.
\end{defn_new}
\vspace{-7pt}
Here, we consider set-valued maps rather than functions to also consider cases in which an input trajectory can have more than one associated output trajectory, e.g., due to initial conditions or non-determinism. 
\vspace{-5pt}
\begin{exam}
Consider the class of systems governed by
\vspace{-15pt}
\begin{align} \label{eq.Systems}
    &x(0) \in \X_0\nonumber\\ 
    &x(k+1) \in \F(x(k),d(k)),~\forall k \in \N\\\nonumber
    &y(k) \in \HH(x(k),d(k)),~\forall k \in \N,
\end{align}

\vspace{-15pt}
where $x \in \Sig^{n_x}$ is the state of the system, $\X_0$ is a set of admissible initial conditions, $\mathcal{F} : \R^{n_x} \times \R^{n_d} \rightrightarrows \R^{n_x}$ is a set-valued map defining the state evolution, and $\mathcal{H} : \R^{n_x} \times \R^{n_d} \rightrightarrows \R^{n_y}$ is a set-valued map defining the observation. This class of systems is included within Definition \ref{def.IOMap}. Moreover, it contains all systems with both linear and non-linear (time-invariant) dynamics, as well as perturbed, unperturbed or uncertain dynamics. Thus, the formalism of Definition \ref{def.IOMap} includes many systems often considered within the scope of control theory.
\end{exam}
Systems governed by \eqref{eq.Systems} are always causal, i.e., the output up to time $k$ is independent of inputs beyond time $k$. Causality will be the key property allowing us to define composition for general networks in Section \ref{sec.GeneralFeedback}. We therefore define the notion of causality for general systems described by Definition \ref{def.IOMap}:

\begin{defn_new} \label{def.Causality}
Let $\Pi : \Sig^{n_d} \rightrightarrows \Sig^{n_y}$ be a system with input $d \in \Sig^{n_d}$ and output $y \in \Sig^{n_y}$. Let $d_1 \in \Sig^{n_{d_1}}$ be a subsignal of $d$.
$\Pi$ is \emph{causal} with respect to $d_1$ if for any time $k$, $y(k)$ does not depend on $d_1(k+1),d_1(k+2),\ldots$. $\Pi$ is \emph{strictly causal} with respect to $d_1$ if for any time $k$, $y(k)$ is also independent of $d_1(k)$. 
If $\Pi$ is causal with respect to $d$, we say it is causal, without mentioning a subsignal.
\end{defn_new}

\vspace{-3pt}
\begin{rem_new} \label{rem.timewise}
Causality with respect to $d$, à la Definition~\ref{def.Causality}, is equivalent to the standard definition of causality using truncation operators [\cite{Desoer2009}]. Equivalently, there is a one-to-one correspondence between causal systems $\Pi : d\mapsto y$ and sets of timewise set-valued maps $\{\Pi_k\}_{k\in \N}$ mapping $d(0:k)$ to $y(k)$. 
\end{rem_new}
 
 \addtocounter{exam}{-1}
 \vspace{-3pt}
\begin{exam}[Continued]
A system $\Pi$ governed by \eqref{eq.Systems} is always causal, and is strictly causal if and only if $\HH$ is independent of $d(k)$.
\end{exam}

We consider specifications on the behaviour of these systems
via the formalism of assume/guarantee contracts. 

\vspace{-3pt}
\begin{defn_new} \label{defn.AG}
An assume/guarantee (A/G) contract is a pair $(\D,\OO)$ where $\D \subseteq \Sig^{n_d}$ are the assumptions and $\OO \subseteq \Sig^{n_d} \times \Sig^{n_y}$ are the guarantees. 
\end{defn_new}
The guarantees on the output $y(\cdot)$ given the input $d(\cdot)$ are manifested by specifications on the input-output pair $(d(\cdot),y(\cdot))$. A/G contracts prescribe specifications on dynamical systems via the notion of satisfaction:

\vspace{-3pt}
\begin{defn_new}
The system $\Pi$ satisfies $\C = (\D,\OO)$ if, for any $d\in \D$ and $y\in \Pi(d)$, we have $(d,y)\in \OO$. In that case, we write $\Pi \sat \C$.
\end{defn_new}
Section \ref{sec.ProbForm} will consider a similar, although different, framework for contracts, which will be more compatible with feedback composition.

One of the main strengths of contract theory is its modularity. Namely, a contract on a composite system can be refined by a collection of ``local" contracts on individual subsystems or components [\cite{Benveniste2018}]. This idea hinges on two notions, refinement and composition. Refinement considers two contracts on the same system, and determines when one is implied  by the other. Composition defines the coupling of multiple contracts on different components. Our goal is to provide such a modular framework for general networks of dynamical control systems. 
\begin{defn_new} \label{def.refine}
Let $\C = (\D,\OO)$ and $\C^\prime = (\D^\prime,\OO^\prime)$ be contracts on the same system with input $d\in \Sig^{n_d}$ and output $y\in \Sig^{n_y}$. We say $\C$ \emph{refines} $\C^\prime$ (and write $\C \preccurlyeq \C^\prime$) if $\D \supseteq \D^\prime$ and $\Omega \cap (\D^\prime \times \Sig^{n_y}) \subseteq \Omega^\prime \cap(\D^\prime \times \Sig^{n_y})$.
\end{defn_new}
Colloquially, $\C \preccurlyeq \C^\prime$ if $\C$ assumes less than $\C^\prime$, but guarantees more. Cascaded contract composition
will be introduced in Section \ref{sec.Feedbackless}.

\if(0)
\subsection{Polyhedral Sets}

The LP-based tools developed for LTI contracts arise from basic facts about polyhedral sets. 
\begin{defn_new}
A set $S \subseteq \R^{d}$ is \emph{polyhedral} if there exist a matrix $A$ and a vector $b$ such that $S = \{z\in \R^d: Az \le b\}$, i.e., $S$ is the intersection of finitely-many half-spaces.
\end{defn_new}
Polyhedral sets are convex. Moreover, optimising a linear function over a polyhedral set results in a linear programme (LP), which can be efficiently solved using existing optimisation software.
Any polyhedral set has an equivalent representation, known as the vertex representation: \begin{lem_new}[\hspace{0.1pt}\cite{Schrijver1998}]
$S \subseteq \R^d$ is a polyhedral set if and only if there exist matrices $F,G$ such that $S = \{F\lambda + G\theta : \mathds{1}^\top \lambda = 1,~\lambda,\theta \ge 0\}$.
\end{lem_new}
Both representations are useful for different purposes, e.g., the vertex representation is useful when computing the Minkowski sum of two polyhedral sets, whereas the subspace representation $\{z: Az \le b\}$ is useful when calculating the preimage under a linear transformation.

Throughout the paper, we repeatedly encounter the inclusion verification problem, i.e., given two polyhedral sets $S_1,S_2$, how can we verify whether $S_1 \subseteq S_2$ holds?
\begin{lem_new}(\hspace{0.1pt}\cite{Sharf2021b}) \label{lem.Inclusion}
Let $S_1,S_2 \subseteq \R^n$ be polyhedral sets.
\begin{itemize}
    \item If $S_r = \{z : A_r z \le b_r\}, r=1,2$ , then $S_1 \subseteq S_2$ if and only if $\varrho_l \le 0$ for any $l$, where $\varrho_l$ is given by the following LP
    \begin{align*}
        \varrho_l = \max\{{\rm e}^\top_l (A_2 z - b_2) : A_1 z \le b_1\}.
    \end{align*}
    \item If $S_r = \{F_r \lambda + G_r \theta : \mathds{1}^\top \lambda = 1,~\lambda,\theta \ge 0\}, r=1.2$, then $S_1 \subseteq S_2$ if and only if there exist matrices $\Lambda,\Theta_F,\Theta_G$ {with non-negative entries} satisfying
    \begin{align*} 
        {G_1 = G_2\Theta_G,~F_1 = F_2\Lambda + G_2\Theta_F,~\Lambda^\top \mathds{1} = \mathds{1}.}
    \end{align*}
\end{itemize}
\end{lem_new}
\fi

\vspace{-8pt}
\subsection{Networked Systems and Graph Theory}
\vspace{-8pt}
The study of networked systems requires an exact description of the interconnection of the different components, which is usually manifested using graph theory. A graph $\G = (\V,\E)$ consists of a set of vertices (or nodes), $\V$, and a set of edges $\E$, which are pairs of vertices. In this paper, we consider \emph{directed} graphs. If $i,j\in \V$, the edge $e$ from $i$ to $j$ is denoted $i \to j \in \E$, and we say that $i$ is $e$'s tail, and $j$ its head. 
A path is a sequence of edges $e_1,e_2,\ldots,e_l \in \E$ such that $e_r$'s head is $e_{r+1}$'s tail for all $r\in\mathcal{I}_{1,l-1}$. The path is called a cycle if $e_l$'s head is $e_1$'s tail. For a node $i\in \V$, the node $j\in \V$ is \emph{backward-reachable} from $i$ if there exists a path from $j$ to $i$. The collection of all backward-reachable nodes from $i\in \V$ is denoted $\BR(i)$. We also denote $\BR_+(i) = \BR(i)\cup\{i\}$.

\begin{figure*}
    \centering
    \subfigure[The graph $\G$, which is a DAG.] {\scalebox{.46}{\includegraphics{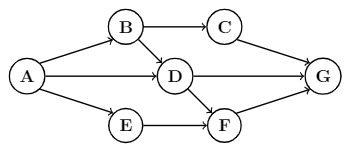}}} \hspace{5pt}
    \subfigure[The node $C$ (red) and $\BR(C)$ (blue)] {\scalebox{.46}{\includegraphics{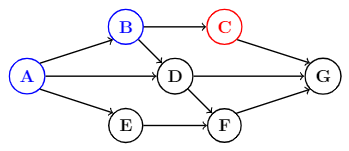}}}\hspace{5pt}
    \subfigure[The node $F$ (red) and $\BR(F)$ (blue)] {\scalebox{.46}{\includegraphics{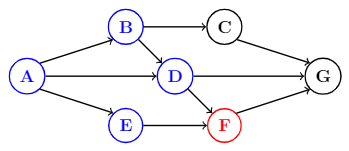}}}
    \vspace{-8pt}
    \caption{An example of a DAG $\G$ and two backward-reachable sets. The graph $\G$ has a total of $11$ different topological orderings, including ABCDEFG, ABDEFCG and AEBDFCG.}
    \label{fig.Graphs}
    \vspace{-4pt}
\end{figure*}

\vspace{-2pt}
A directed acyclic graph (DAG) is a directed graph $\G$ containing no cycles. DAGs play a vital role in algorithm design and analysis as many problems, e.g., the shortest-path and the longest-path problems, are solvable in linear-time on these graphs. On general directed graphs, however, the former requires more time, and the latter is \textbf{NP}-hard [\cite{Cormen2009}]. This acceleration hinges on the tool of topological ordering:

\vspace{-3pt}
\begin{defn_new} \label{def.TopoSort}
Let $\G = (\V,\E)$ be a graph with $N$ nodes. A topological ordering is a map $\sigma:\mathcal{I}_{1,N} \to \V$ such that: 
\vspace{-8pt}
\begin{itemize}
    \item[i)] If $p, q \in \mathcal{I}_{1,N}$ satisfy $p\neq q$, then $\sigma(p)\neq \sigma(q)$.
    \item[ii)] If $p,q\in \mathcal{I}_{1,N}$ satisfy $\sigma(p)\to \sigma(q) \in \E$, then $p < q$.
\end{itemize}
\end{defn_new}
Occasionally topological orderings are given as a list rather than a function. For example, the list $v_1v_2v_3v_4v_5$ corresponds to the function $\sigma : \mathcal{I}_{1,5}\to \V$ defined by $\sigma(1)=v_1$, $\sigma(2) = v_2$, etc. An example of a DAG $\G$, together with some for the sets $\BR(i)$ and topological orderings, can be found in Fig. \ref{fig.Graphs}.

Pictorially, a topological ordering is an ordering of the vertices on a horizontal line such that all edges go from left to right.
A graph has a topological ordering if and only if it is a DAG. There are linear-time algorithms for finding a topological ordering of a DAG, and for checking whether a graph is a DAG, e.g., relying on depth-first search [\cite{Cormen2009}, p. 613-614]. 
We will repeatedly apply the following lemma connecting backward-reachability and topological ordering.

\vspace{-3pt}
\begin{lem_new} \label{lem.DAGSort}
Let $\G = (\V,\E)$ be a DAG with topological ordering $\sigma : \mathcal{I}_{1,N} \to \V$. For any $q\in \mathcal{I}_{1,N}$, we have that $\BR(\sigma(q)) \subseteq \{\sigma(1),\ldots,\sigma(q-1)\}$.
\end{lem_new}
\vspace{-15pt}
\begin{pf}
Follows from the part ii) of Definition \ref{def.TopoSort}. \qedwhite
\end{pf}
\vspace{-15pt}

We attach a graph $\G = (\V,\E)$ to each networked dynamical system by fitting each component $C_i$ with a vertex $i$, and inserting an edge $i\to j$ if the output of $C_i$ is used as an input to $C_j$. In other words, thinking of the networked system as a block diagram, the corresponding graph $\G$ is achieved by treating the blocks as vertices and the lines between them as edges.\footnote{We omit all exogenous inputs and outputs, i.e., "lines" in the block diagram touching only one block.} Thus, feedback loops in the networked system correspond to cycles in the graph, i.e., the networked system is feedback-less if and only if the associated graph is a DAG.

\vspace{-9pt}
\section{Problem Formulation} \label{sec.ProbForm}
\vspace{-9pt}
This section presents the problem formulation. It first extends the definition of contracts to be compatible with feedback control, and then states the requirements on contract composition and vertical contracts.

\vspace{-6pt}
\subsection{Generalised Causal Contracts}
\vspace{-8pt}
The definition of contracts presented in \cite{Sharf2021b}, stemming from \cite{Benveniste2018}, prescribes assumptions on the input signal $d(\cdot)$ and guarantees on the output signal $y(\cdot)$. This approach is intuitive when adapting the abstract theory of \cite{Benveniste2018} to control systems, and is applicable in various scenarios. Unfortunately, it is a bit restrictive for dynamical control systems. 
For example, consider a vehicle with the control input $d(\cdot)$, and the output $y(\cdot)$ equal to the velocity of the vehicle. If we wish to guarantee that the velocity of the vehicle is below some limit $V_{\rm max}$, the set of admissible values for $d(k)$ must depend on the velocity. However, this assumption cannot be accommodated in the existing framework, as it restricts $d(\cdot)$ in terms of $y(\cdot)$. 
We remedy the problem by considering a more general class of specifications, allowing the assumptions at any time to depend on previous outputs.

\vspace{-3pt}
\begin{defn_new} \label{def.GCAG}
A \emph{recursively-defined} (RD) \emph{contract} is a pair $(\D,\OO)$ of sets inside $\Sig^{n_d+n_y}$, where $\D$ are the assumptions and $\Omega$ are the guarantees. Moreover, we have
\begin{align} \label{eq.GCAG_A}
    \D &= \left\{\left(\begin{smallmatrix}d(\cdot)\\y(\cdot)\end{smallmatrix}\right): d(k) \in A_k\hspace{-2.5pt}\left(\begin{smallmatrix} d(0:k-1)\\y(0:k-1)\end{smallmatrix}\right),\forall k\right\}, \\
    \label{eq.GCAG_G}
    \OO &= \left\{\left(\begin{smallmatrix}d(\cdot)\\y(\cdot)\end{smallmatrix}\right): \left[\begin{smallmatrix}d(k) \\ y(k)\end{smallmatrix}\right] \in G_k\hspace{-2.5pt}\left(\begin{smallmatrix} d(0:k-1)\\y(0:k-1)\end{smallmatrix}\right),\forall k\right\},
\end{align}
for some set-valued functions $A_k: (\R^{n_d} \times \R^{n_y})^k \rightrightarrows \R^{n_d}$ and $G_k: (\R^{n_d} \times \R^{n_y})^k \rightrightarrows \R^{n_d} \times \R^{n_y}$.
\end{defn_new}
In other words, RD contracts put assumptions on the input $d(k)$ in terms of the previous inputs and outputs, and guarantees on the output $y(k)$ in terms of the previous inputs, the previous outputs, and the current input. We emphasise that although the assumptions are written as $(d,y)\in \D$, they only restrict $d(k)$. The assumptions are allowed to ``react" to $y(0:k-1)$, but cannot restrict it in any form.

\vspace{-5pt}
\begin{exam}
Consider a dynamical system with input $d(\cdot) \in \Sig^2$ and output $y(\cdot) \in \Sig^1$. The input $d(\cdot)$ has two subsignals $d_1,d_2$. The signal $d_1(\cdot) \in \Sig^1$ is a disturbance that should be rejected, and the signal $d_2(\cdot) \in \Sig^1$ is a control input. We assume $d_1(\cdot)$ is small, and that $d_2(\cdot)$ is the output of a proportional controller with gain $K$ and a small actuation error. We wish to guarantee that $y(\cdot)$ is close enough to zero. This specification can be expressed as the following RD contract $\C = (\D,\OO)$:

\vspace{-25pt}
\begin{align*}
    \D &= \{(d,y): |d_1(k)|\le \epsilon_1,|d_2(k) - Ky(k-1)| \le \epsilon_2,\forall k\},\\
    \OO &= \{(d,y): |y(k)| \le \epsilon_3,\forall k\}
\end{align*}
\end{exam}

\vspace{-15pt}
\begin{rem_new}
Not all A/G contracts are also RD contracts, e.g., the assumption $\D = \{d(\cdot): \sum_{k=0}^\infty |d(k)|^2 \le 1\}$ can be included in an A/G contract, but not in an RD contract. However, A/G contracts defined by time-invariant inequalities are also RD contracts. 
Moreover, we could consider more general definitions of RD contracts. For example, the demand \eqref{eq.GCAG_G} will only be required when considering networks with feedback (see Section \ref{sec.GeneralFeedback}). One could also consider assumptions of the form $\D = (\D_0 \times \Sig^{n_y}) \cap \D_y$ where $\D_0 \subset \Sig^{n_d}$ and $\D_y$ is of the form \eqref{eq.GCAG_A}. This extended definition includes all A/G contracts, and later theorems still hold with exactly the same proof. However, we restrict ourselves and use Definition~\ref{def.GCAG} for simplicity of the presentation.
\end{rem_new}

The definitions of satisfaction and refinement must be adapted accordingly to accommodate RD contracts:

\vspace{-3pt}
\begin{defn_new}
Let $\Pi$ be a system and $\C = (\D,\OO)$ be an RD contract with input $d\in \Sig^{n_d}$ and output $y\in \Sig^{n_y}$. We say $\Pi$ satisfies $\C$ (and write $\Pi \sat \C$) if for any $d\in \Sig^{n_d},y \in \Sig^{n_y}$, if $y \in \Pi(d)$ and $(d,y) \in \D$ hold, then $(d,y)\in \OO$.
\end{defn_new}

\vspace{-3pt}
\begin{defn_new}
Let $\C = (\D,\OO)$ and $\C^\prime = (\D^\prime,\OO^\prime)$ be two RD contracts on the same system. We say that $\C$ refines $\C^\prime$ (and write $\C \preccurlyeq \C^\prime$) if $\D \supseteq \D^\prime$ and $\Omega \cap \D^\prime \subseteq \Omega^\prime \cap \D^\prime$.
\end{defn_new}

For the LP-tools developed later in this paper, we make the following assumption:

\vspace{-3pt}
\begin{defn_new}\label{def.LinConGCAG}
A \emph{linear time-invariant} (LTI) RD contract $\C = (\D,\OO)$ of assumption depth $m_A\in \N$ and guarantee depth $m_G \in \N$ is given by matrices $\{\mathfrak A^r\}_{r=0}^{m_A}$, $\{\mathfrak G^r\}_{r=0}^{m_G}$ and vectors $\mathfrak{a}^0,\mathfrak{g}^0$ of appropriate sizes, where:

\textcolor{white}{.} 

\textcolor{white}{.} 
\vspace{-36pt}
\small
\begin{align} \label{eq.LTICont} \nonumber
    \D &= \left\{\left(\begin{smallmatrix}d \\ y \end{smallmatrix}\right): \mathfrak A^0 d(k) + \sum_{r=1}^{m_A} \mathfrak A^r \left[\begin{smallmatrix}d(k-m_A+r) \\ y(k-m_A+r) \end{smallmatrix}\right] \le \mathfrak a^0,~\forall k\ge m_A\right\}\\ 
    \OO &= \left\{\left(\begin{smallmatrix}d \\ y \end{smallmatrix}\right): \sum_{r=0}^{m_G} \mathfrak G^r \left[\begin{smallmatrix} d(k-m_G+r) \\ y(k-m_G+r) \end{smallmatrix}\right] \le \mathfrak g^0,~ \forall k\ge m_G\right\}.
\end{align}\normalsize
\end{defn_new}

\vspace{-8pt}
\begin{rem_new}
We may assume that $m_A,m_G\ge 1$, as contracts of depth $0$ are also contracts of depth $1$. 
\end{rem_new}

\vspace{-6pt}
For any LTI RD contract of the form \eqref{eq.LTICont}, we consider two associated piecewise-linear functions $\alpha:(\R^{n_d})^{m_A+1} \times (\R^{n_y})^{m_A} \to \R$ and $\gamma:(\R^{n_d+n_y})^{m_G+1} \to \R$, given by $\alpha = \alpha \left(\begin{smallmatrix}d(0:m_A) 
\\ y(0:m_A-1) \end{smallmatrix}\right)$ and $\gamma = \gamma \left(\begin{smallmatrix}d(0:m_G) 
\\ y(0:m_G) \end{smallmatrix}\right)$, and defined as
    
\vspace{-22pt}
\begin{align} \label{eq.ConvLTIFunc} \nonumber
    &\alpha = \max_{i} {\rm e}_i^\top \left(\mathfrak A^0 d(k) + \sum_{r=1}^{m_A} \mathfrak A^r \left[\begin{smallmatrix}d(k-m_A+r) \\ y(k-m_A+r) \end{smallmatrix}\right] - \mathfrak a^0\right)\\
    &\gamma = \max_{i} {\rm e}_i^\top \left(\sum_{r=0}^{m_G}\mathfrak G^r \left[\begin{smallmatrix} d(k-m_G+r) \\ y(k-m_G+r) \end{smallmatrix}\right] - \mathfrak g^0\right),
\end{align}

\vspace{-20pt}
Thus, the contract \eqref{eq.LTICont} can be written as:

\vspace{-23pt}
\begin{align*}
    \D &= \left\{\left(\begin{smallmatrix}d \\ y \end{smallmatrix}\right): \alpha\left( \begin{smallmatrix}d(k-m_A:k) \\ y(k-m_A:k-1) \end{smallmatrix}\right) \le 0,~\forall k\ge m_A\right\}\\ 
    \OO &= \left\{\left(\begin{smallmatrix}d \\ y \end{smallmatrix}\right): \gamma\left( \begin{smallmatrix}d(k-m_G:k) \\ y(k-m_G:k) \end{smallmatrix}\right) \le 0,~ \forall k\ge m_G\right\}.
\end{align*}

\vspace{-15pt}
\begin{rem_new}
If $\alpha_1,\alpha_2$ are the piecewise-linear functions associated with such specifications, 
then the maximum $\alpha_{\rm max} = \max\{\alpha_1,\alpha_2\}$ is the piecewise linear function associated with the conjunction of the specifications.
\end{rem_new}

\vspace{-3pt}
Lastly, let us define the notion of extendibility converting assumptions on $d(\cdot)$ by assumptions on $d(0:n)$ for times $n\in \N$. It manifests the self-consistency of the set of assumptions, in the sense that a signal satisfying the assumptions up to time $k$ can be extended beyond time $k$ while still satisfying the assumptions. While the notion was originally defined for A/G contracts in \cite{SharfADHS2020}, we extend it for RD contracts:

\vspace{-3pt}
\begin{defn_new}
Let $\D \subseteq \Sig^{n_d} \times \Sig^{n_y}$ be a set of the form \eqref{eq.GCAG_A}. The set $\mathcal{D}$ is \emph{extendable} if the following condition holds for any $k\in \N$ and any signals $d(\cdot),y(\cdot)$ defined at times $\{0,\ldots,k\}$. If $d(\ell+1) \in A_\ell\left(\begin{smallmatrix}d(0:\ell)\\y(0:\ell)\end{smallmatrix}\right)$ holds for all $\ell \in  \mathcal{I}_{0,k-1}$, then the set $A_{k}\left(\begin{smallmatrix}d(0:k-1)\\y(0:k-1)\end{smallmatrix}\right)$ is non-empty.
\end{defn_new}

\vspace{-3pt}
\begin{rem_new} \label{rem.ExtendableAlpha}
For LTI contracts, we abuse the notation and say that $\alpha$ is extendable if $\D$ is, where \eqref{eq.ConvLTIFunc} holds.
\end{rem_new}

\vspace{-5pt}
\subsection{Contract Composition and Vertical Contracts}
\vspace{-10pt}
Consider a networked system with multiple components, having an associated graph $\G = (\V,\E)$, where each component $i\in \V$ is fitted with a contract $\C_i = (\D_i,\OO_i)$ with input $d_i \in \Sig^{n_{d_i}}$ and output $y_i \in \Sig^{n_{y_i}}$. The input to the $i$-th component, $d_i$, is composed of an external input, $d_i^{\rm ext} \in \Sig^{n_{d_i^{\rm ext}}}$, and the output of some of the other agents, $\{y_j\}_{j\to i\in \E}$, i.e., we have:

\vspace{-27pt}
\begin{align*}
    d_i(k) = [d_i^{\rm ext}(k), y_{j_1}(k),\ldots,y_{j_l}(k)],~\forall k\in \N,
\end{align*}

\vspace{-18pt}
where $j_1,\ldots,j_l$ is a list of the nodes $j$ with $j\to i \in \E$. We introduce matrices $F_{ij} \in \R^{n_{d_i}\times n_{y_j}}$ and $E_i \in \R^{n_{d_i} \times n_{d_i^{\rm ext}}}$ for $i,j\in \V$ such that for any $k\in \N$ and any $i\in \V$,

\vspace{-28pt}
\begin{align} \label{eq.diDefinition}
    d_i(k) = {\textstyle \sum}_{j\in \V} F_{ij}y_j(k) + E_{i} d_i^{\rm ext}(k).
\end{align}

\vspace{-18pt}
Our first goal in this paper is to define the composition of the ``local" contracts $\{\C_i\}_{i \in \V}$, which should be a contract on the composite system. The input to this composite system would be $d^{\rm ext}$, i.e., the signal created by stacking $\{d_i^{\rm ext}\}_{i\in\V}$, but there is no clear candidate for its output. We therefore choose a set $\W = \{w_1,\ldots,w_M\} \subseteq \V$ of ``output components", and define the output as $y^{\rm ext} = [y_{w_1},\ldots,y_{w_M}]$. As before, we find matrices $\{H_i\}_{i\in \V}$ such that

\vspace{-28pt}
\begin{align} \label{eq.OutputConsistency}
y^{\rm ext} (k) = {\textstyle \sum}_{i\in \V} H_i y_i(k),
\end{align}
\vspace{-30pt}

Stating the requirements on contract composition requires us to first define system composition:

\vspace{-5pt}
\begin{defn_new} \label{defn.SysComposition}
Consider a graph $\G = (\V,\E)$, systems $\{\Pi_i\}_{i\in \V}$ at each node and a set $\W \subseteq \V$ of output nodes. The system $\Pi_i$ has input $d_i \in \Sig^{n_{d_i}}$ and output $y_i \in \Sig^{n_{y_i}}$. The composition $\bigotimes_{i\in \V}^{\G,\W} \Pi_i$ is a system with input $d^{\rm ext}$ and output $y^{\rm ext} $, defined by the following set-valued function. We say that $y^{\rm ext}  \in \bigotimes_{i\in \V}^{\G,\W} \Pi_i(d^{\rm ext})$ if there exist signals $d_i \in \Sig^{n_{d_i}}$ and $y_i \in \Sig^{n_{y_i}}$ such that the following consistency relations hold:

\vspace{-27pt}
\begin{align} \label{eq.SysComposition}
        &y_i \in \Pi_i(d_i) && \forall i\in \V,\\ \nonumber
        &d_i(k) = {\textstyle \sum}_{j\in \V} F_{ij} y_j(k) + E_i d_i^{\rm ext}(k) && \forall i\in \V, k\in \N\\ \nonumber
        &y^{\rm ext} (k) = {\textstyle \sum}_{i\in \V} H_i y_i(k).
\end{align}
When $\G$ and $\W$ are clear from the context, we omit them from the notation and write $\bigotimes_{i\in \V} \Pi_i$.
\vspace{-10pt}
\end{defn_new}
 Definition \ref{defn.SysComposition} states composition in terms of consistency equations, which can be made concrete for instance for systems of the form \eqref{eq.Systems}. However, the definitino also obfuscates the problem of algebraic loops, which might exist even when only considering causal systems. More precisely, any algebraic loop corresponds to a cycle in $\G$ traversing through the nodes $i_1,\ldots,i_l$, where the corresponding systems $\Pi_{i_1},\ldots,\Pi_{i_l}$ are causal (but not strictly causal). A thorough investigation of algebraic loops will be considered in Section \ref{sec.GeneralFeedback}, in which networks with feedback will be considered.

Contract composition is considered by the meta-theory of \cite{Benveniste2018} for abstract contracts, relying on two modularity principles. Namely, given a collection of abstract contracts $\{\C_i\}_{i\in \V}$, the contract composition $\bigotimes_{i\in \V} \C_i$ is defined to satisfy the two following postulates:

\vspace{-5pt}
\begin{itemize}
    \item[A)] Its guarantees are the conjunction of the guarantees of all the $\C_i$-s.
    \item[B)] Its assumptions are defined as the largest set with the following property: for any $i$, the conjugation of these assumptions with the guarantees of $\C_j$ for all $j\neq i$ imply the assumptions of $\C_i$.
\end{itemize}

\vspace{-5pt}
This definition supports modular design. Namely, \cite{Benveniste2018} show that if components $\{\Sigma_i\}_{i\in \V}$ satisfy $\C_i$ for $i\in \V$, then the composite system $\bigotimes_{i\in \V} \Sigma_i$ satisfies the composite contract $\bigotimes_{i\in \V} \C_i$. 

\vspace{-5pt}
Unfortunately, this meta-theoretical definition cannot be directly applied to RD (or even A/G) contracts for dynamical control systems for two main reasons. First, the definition appearing in \cite{Benveniste2018} makes no distinction between external and internal variables, leading to situations in which the set of assumptions for the composed contract refers to the value of internal variables. Similarly, composition is only defined when the network output $y^{\rm ext}$ is composed of \emph{all} ``local" outputs $y_i$. Second, \cite{Benveniste2018} does not propose any computational tools for composition, e.g., a way to verify that a given contract on a network system in refined by the composition of component-level contracts. The goal of this paper is to address both of these problems, specifically for contracts on (causal) dynamical control systems. This goal is explicitly formulated in the following problem statements:

\vspace{-5pt}
\begin{prob_new}\label{prob.1}
Given a graph $\G = (\V,\E)$, RD contracts $\{\C_i\}_{i\in \V}$, and a set $\W \subseteq \V$ of output nodes, define the composite contract $\bigotimes_{i\in \V}^{\G,\W} \C_i$, with input $d^{\rm ext}$ and output $y^{\rm ext}$ in a way compatible with postulates A) and B), while only using the external input $d^{\rm ext}$ and output $y^{\rm ext}$.
\end{prob_new}

\vspace{-5pt}
We also show our definition satisfies the universal property of composition, namely, that if $\Pi_i$ are causal systems with $\Pi_i \sat \C_i$ for $i\in \V$, then $\bigotimes_{i\in \V} \Pi_i \sat \bigotimes_{i\in \V} \C_i$.

\vspace{-5pt}
Once composition is defined, we can address the connection between contracts on different levels of abstraction:

\vspace{-5pt}
\begin{defn_new}
Consider a networked system with a graph $\G = (\V,\E)$ and a set $\W \subseteq \V$ of output nodes. A vertical contract is a statement of the form $\bigotimes_{i\in\V} \C_i \preccurlyeq \C_{\rm tot}$, with $\C_{\rm tot}$ an RD contract on the composite networked system and  $\{\C_i\}_{i\in \V}$ are component-level RD contracts.
\end{defn_new}

\vspace{-5pt}
\begin{prob_new} \label{prob.2}
Find a computationally viable algorithm checking if a vertical contract $\bigotimes_{i\in\V} \C_i \preccurlyeq \C_{\rm tot}$ holds. 
\end{prob_new}

\vspace{-5pt}
The main strength of contract theory hinges on solving Problems \ref{prob.1} and \ref{prob.2}. Indeed, we prove modularity-in-design is achieved:

\vspace{-5pt}
\begin{thm_new}
Consider a graph $\G = (\V,\E)$, component-level RD contracts $\{\C_i\}_{i\in \V}$ and an output set $\W \subseteq \V$. Let $\C_{\rm tot}$ be an RD contract on the composite system, where the composition $\bigotimes_{i\in \V} \C_i$ is defined, and the vertical contract $\bigotimes_{i\in \V} \C_i \preccurlyeq \C_{\rm tot}$ holds.
If the systems $\{\Pi_i\}_{i\in \V}$ satisfy $\Pi_i \sat \C_i$ for all $i\in \V$, then $\bigotimes_{i\in \V} \Pi_i \sat \C_{\rm tot}$.
\end{thm_new}

\vspace{-15pt}
\begin{pf}
Follows directly from Proposition 1 of \cite{SharfADHS2020}, as we have $\bigotimes_{i\in \V} \Pi_i \sat \bigotimes_{i\in \V} \C_i \preccurlyeq \C_{\rm tot}$. \qed
\end{pf}

\vspace{-21pt}
Before moving on to the solutions to these problems first for feedback-less networks and later for general networks, we make an important remark about the output set $\W$. RD contracts allow the assumption to depend on previous outputs, and these assumptions should still be manifested in the composition. Thus, relevant ``local" outputs $y_i$ must be available as a part of the ``global" output $y^{\rm ext}$:

\vspace{-5pt}
\begin{assump} \label{assump.1}
For any $i\in \V$, if the assumption on the external input $d^{\rm ext}_i$ explicitly depends on the output $y_i$ in the RD contract $\C_i$, then $i\in \W$.
\end{assump}
\vspace{-7pt}

In other words, if the component-level assumption on the external input depends on the output $y_i$, then $y_i$ should be a part of the output of the composite system. This assures that the assumptions of the composition will not depend on any internal variables.

\vspace{-10pt}
\section{Networks Without Feedback} \label{sec.Feedbackless}
\vspace{-10pt}
In this section, we propose solutions to Problems \ref{prob.1} and \ref{prob.2} for feedback-less networked systems, e.g., networks with open-loop control. We first define composition for networks without feedback, and then show that the correctness of vertical contracts can be verified using LP-enabled tools.
For this section, fix a networked control system with the underlying graph $\G = (\V,\E)$, assumed to be a DAG, component-level RD contracts $\{\C_i\}_{i\in \V}$, and a subset $\W \subseteq \V$ of output components, so that Assumption \ref{assump.1} holds. 

\vspace{-6pt}
\subsection{Defining Composition}
\vspace{-10pt}
We wish to define the composite contract $\bigotimes_{i\in\V} \C_i$ as to satisfy postulates A) and B), while only using the external input $d^{\rm ext}$ and the external output $y^{\rm ext}$. Postulate A), defining the guarantees of the composition, will be adapted by requiring the existence of signals $(d_i,y_i)\in \OO_i$ for $i\in \V$ such that the consistency conditions \eqref{eq.diDefinition} and \eqref{eq.OutputConsistency} hold. As for postulate B), instead of considering all components $j\neq i$, it suffices to consider components $j$ which precede $i$ (in the sense of backward reachability). Indeed, these are the only components whose output can affect input $d_i$, as there are no feedback loops.

\vspace{-5pt}
\begin{defn_new} \label{defn.Feedbackless}
Let $\G = (\V,\E)$ be a feedback-less network, $\W\subseteq\V$ be a set of output nodes, and $\C_i = (\D_i,\Omega_i)$ be component-level RD contracts, so that Assumption \ref{assump.1} holds. The composition $\bigotimes_{i\in \V} \C_i = (\D_\otimes,\Omega_\otimes)$, having input $d^{\rm ext}(\cdot)$ and output $y^{\rm ext}(\cdot)$, is defined as follows:
\begin{itemize}
    \item $(d^{\rm ext},y^{\rm ext}) \in \D_\otimes$ if for any signals $\{d_i(\cdot),y_i(\cdot)\}_{i\in \V}$ satisfying the input-consistency constraints \eqref{eq.diDefinition} and output-consistency constraints \eqref{eq.OutputConsistency}, the following implication holds for all $i\in \V$: if $(d_j,y_j) \in \Omega_j$ holds for any $j \in \BR(i)$, then $(d_i,y_i)\in \D_i$. 
    \item $(d^{\rm ext},y^{\rm ext} ) \in \Omega_\otimes$ if there are signals $\{d_i(\cdot),y_i(\cdot)\}_{i\in \V}$ such that $(d_i,y_i) \in \Omega_i$ holds for $i \in \V$, and the input- and output-consistency constraints \eqref{eq.diDefinition} and \eqref{eq.OutputConsistency} hold.
\end{itemize}
\vspace{-5pt}
\end{defn_new}

It can be shown that this composition of RD contracts is itself an RD contract, and in particular, the input signal $y^{\rm ext}$ is a free variable in $\D_\otimes$. 
We next prove that the universal property of composition is satisfied:

\vspace{-5pt}
\begin{thm_new} \label{thm.CompFeedbackless}
Let $\G = (\V,\E)$ be a feedback-less network, with component-level RD contracts $\C_i = (\D_i,\Omega_i)$, and an output set $\W \subseteq \V$. Let $\{\Pi_i\}_{i\in\V}$ be systems. If $\Pi_i \sat \C_i$ for all $i\in \V$, then $\bigotimes_{i\in \V} \Pi_i \sat \bigotimes_{i\in\V} \C_i$.
\end{thm_new}

\vspace{-15pt}
\begin{pf}
We must show that if $(d^{\rm ext},y^{\rm ext}) \in \D_\otimes$ and $y^{\rm ext} \in \bigotimes_{i\in \V} \Pi_i (d^{\rm ext})$ both hold, then $(d^{\rm ext},y^{\rm ext}) \in \OO_\otimes$. As the network was assumed to be feedback-less, the graph $\G$ is a DAG. We can thus find a topological ordering $\sigma:\mathcal{I}_{1,N} \to \V$ of $\G$ satisfying Definition \ref{def.TopoSort}.
By Definition \ref{defn.SysComposition}, there exist signals
$\{d_i\}_{i\in \V}$ and $\{y_i\}_{i\in \V}$ such that \eqref{eq.SysComposition} holds. We prove that $(d_i,y_i) \in \Omega_i$ holds for $i\in \V$, implying that $(d^{\rm ext},y^{\rm ext}) \in \OO_\otimes$. We do so by writing $i=\sigma(q)$ for $q\in\mathcal{I}_{1,N}$ and using induction on $q$.

We first consider the basis $i=\sigma(1)$. By Lemma \ref{lem.DAGSort}, $\BR(i) = \emptyset$. Thus, by the definition of the matrices $F_{ij}$, we have that $d_i = d_i^{\rm ext}$, and the assumption that $(d^{\rm ext},y^{\rm ext} ) \in \D_\otimes$ together with Definition \ref{defn.Feedbackless} imply that $(d_i,y_i) = (d_i^{\rm ext},y_i) \in \D_i$. Hence, $(d_i,y_i)\in \OO_i$ as $y_i \in \Pi_i(d_i)$ and $\Pi_i \sat \C_i$.
For the induction step, we write $i=\sigma(q)$ and assume $(d_j,y_j) \in \OO_j$ holds for all $j = \sigma(p)$ for $p \in \mathcal{I}_{1,q-1}$. In particular, $(d_j,y_j) \in \OO_j$ holds for any $j\in \BR(i)$ by Lemma \ref{lem.DAGSort}. As $(d^{\rm ext},y^{\rm ext} ) \in \D_\otimes$, we conclude that $(d_i,y_i) \in \D_i$ by Definition \ref{defn.Feedbackless}. We therefore see that $(d_i,y_i) \in \Omega_i$ using $y_i \in \Pi_i(d_i)$, as $\Pi_i \sat \C_i$.  \qed
\end{pf}

\vspace{-15pt}
\begin{rem_new}
Definition \ref{defn.Feedbackless} is stated for RD contracts, and serves as a stepping stone for defining contract composition for general networks with feedback. An almost identical definition can be made for A/G contracts\footnote{Replace statements of the form $(d,y)\in \D$ with $d\in \D$.}, for which a similar result to Theorem \ref{thm.CompFeedbackless} still holds.
\end{rem_new}

\begin{rem_new}
Definition \ref{defn.Feedbackless} considers the assumptions of the composite contract as pairs $(d^{\rm ext},y^{\rm ext})$ satisfying a certain implication. If no such pair exist, so that $\D_\otimes = \emptyset$, one might say that the contracts are \emph{incompatible}, using the terminology of \cite{Benveniste2018}. One example of such case is when a certain contract guarantees that some signal $v$ has $|v(k)|\le 2$, but another contract assumes that $|v(k)| \le 1$, i.e., the guarantee of the former is not strict enough for the latter.
\end{rem_new}

\vspace{-6pt}
\subsection{Vertical Contracts}
\vspace{-10pt}
We now consider Problem \ref{prob.2} for feedback-less networks. We build LP-based tools for verifying vertical contracts of the form $\bigotimes_{i\in \V} \C_i \preccurlyeq \C_{\rm tot}$ for LTI RD contracts. Let $\C_i = (\D_i,\OO_i)$ for $i\in \V$ be component-level LTI RD contracts, and let $\C_{\rm tot} = (\D_{\rm tot},\OO_{\rm tot})$ be an LTI RD contract on the composite system. Assume $\C_i,\C_{\rm tot}$ have assumption depth $m_i^A,m_{\rm tot}^A$ and guarantee depth $m_i^G,m_{\rm tot}^G$, respectively. Denoting the associated piecewise-linear functions as $\alpha_i,\alpha_{\rm tot},\gamma_i,\gamma_{\rm tot}$, we write:

\vspace{-23pt}
\small
\begin{align} \label{eq.NContractsCascade}
    \D_i &= \left\{\left(\begin{smallmatrix}d_i\\y_i\end{smallmatrix}\right): \nonumber \alpha_i\left(\begin{smallmatrix}d_i(k-m^A_i:k) \\ y_i(k-m^A_i:k-1) \end{smallmatrix}\right) \le 0,~\forall k\ge m^A_i\right\},\\ 
    \OO_i &= \left\{\left(\begin{smallmatrix}d_i\\y_i\end{smallmatrix}\right):\gamma_i\left(\begin{smallmatrix} d_i(k-m_i^G:k) \\ y_i(k-m_i^G:k) \end{smallmatrix}\right) \le 0,~\forall k\ge m_i^G\right\},\\\nonumber
    \D_{\rm tot} &= \left\{\left(\begin{smallmatrix}d^{\rm ext}\\y^{\rm ext} \end{smallmatrix}\right): \alpha_{\rm tot}\left(\begin{smallmatrix}d^{\rm ext}(k-m^A_{\rm tot}:k)\\ y^{\rm ext} (k-m_{\rm tot}^A : k-1)\end{smallmatrix}\right) \le 0,~\forall k\ge m_{\rm tot}^A\right\},\\\nonumber
    \OO_{\rm tot} &= \left\{\left(\begin{smallmatrix}d^{\rm ext}\\y^{\rm ext} \end{smallmatrix}\right): \gamma_{\rm tot}\left(\begin{smallmatrix}d^{\rm ext}(k-m_{\rm tot}^G:k)\\ y^{\rm ext} (k-m_{\rm tot}^G : k)\end{smallmatrix}\right) \le 0,~\forall k\ge m_{\rm tot}^G\right\}.\nonumber
\end{align} \normalsize

\vspace{-17pt}
We denote $\bigotimes_{i\in\V} \C_i = (\D_\otimes,\OO_\otimes)$. Our goal is to find a computationally-viable method for verifying that $\bigotimes_{i\in\V} \C_i \preccurlyeq \C_{\rm tot}$ holds. The vertical contract is equivalent to the set inclusions $\D_\otimes \supseteq \D_{\rm tot}$ and $\OO_\otimes \cap \D_{\rm tot} \subseteq \OO_{\rm tot} \cap \D_{\rm tot}$, which can be rewritten as the following implications for the signals $d^{\rm ext}, y^{\rm ext} , \{d_j,y_j\}_{j\in \V}$ satisfying the consistency conditions \eqref{eq.diDefinition}, \eqref{eq.OutputConsistency}:

\vspace{-10pt}
\begin{itemize}
    \item Given any $i \in \V$, if $(d^{\rm ext}(\cdot),y^{\rm ext} (\cdot)) \in \D_{\rm tot}$ and $(d_j(\cdot),y_j(\cdot)) \in \OO_j$ hold for all $j\in \BR(i)$, then $(d_i,y_i) \in \D_i$.
    \item If $(d^{\rm ext}(\cdot),y^{\rm ext} (\cdot)) \in \D_{\rm tot}$ and $(d_i(\cdot),y_i(\cdot)) \in \OO_i$ hold for all $i\in \V$, then $(d^{\rm ext}(\cdot),y^{\rm ext} (\cdot)) \in \OO_{\rm tot}$.
\end{itemize}
\vspace{-6pt}

By using extendibility, we reformulate these as implications on signals defined on a bounded time interval.

\vspace{-10pt}
\begin{thm_new} \label{thm.VertCascadeN}
Consider a feedback-less networked system with DAG $\G = (\V,\E)$ and output set $\W$. Let $\{\C_i\}_{i\in \V},\C_{\rm tot}$ be RD contracts as in \eqref{eq.NContractsCascade}, where Assumption \ref{assump.1} holds. Under mild extendibility assumptions,\footnote{\label{footnote.extend}The functions $\max\{\max_{\ell \le m_i^A} \alpha_{\rm tot}\left(\begin{smallmatrix}d^{\rm ext}(\ell-m_{\rm tot}^A:\ell)\\y^{\rm ext} (\ell-m_{\rm tot}^A:\ell-1)\end{smallmatrix}\right),$ $\max_{j\in \BR(i)}\max_{\ell \le m_i^A} \gamma_j\left(\begin{smallmatrix}d_j(\ell-m_j^G:\ell) \\ y_j(\ell-m_j^G:\ell)\end{smallmatrix}\right)\}$ for $i\in \V$, as well as the function $\max\{\max_{\ell \le m_{\rm tot}^G} \alpha_{\rm tot}\left(\begin{smallmatrix}d^{\rm ext}(\ell-m_{\rm tot}^A:\ell)\\y^{\rm ext} (\ell-m_{\rm tot}^A:\ell-1)\end{smallmatrix}\right),$ $\max_{i\in \V}\max_{\ell \le m^G_{\rm tot}}\gamma_j\left(\begin{smallmatrix}d_j(\ell-m_j^G:\ell) \\ y_j(\ell-m_j^G:\ell)\end{smallmatrix}\right)\}$, are extendable.} $\bigotimes_{i\in \V} \C_i \preccurlyeq \C_{\rm tot}$ holds if and only if the following implications hold for any signals $d_i,y_i,d^{\rm ext}, y^{\rm ext}$ satisfying all input- and output-consistency conditions \eqref{eq.diDefinition},\eqref{eq.OutputConsistency}:

\vspace{-10pt}
\begin{enumerate}
    \item[i)] For any $i\in \V$, if 
    
    \vspace{-15pt}
    \small
    \begin{align*}
    &\alpha_{\rm tot}\left(\begin{smallmatrix}d^{\rm ext}(\ell-m_{\rm tot}^A:\ell) \\ y^{\rm ext} (\ell-m_{\rm tot}^A: \ell - 1)\end{smallmatrix}\right) \le 0,&&\forall\ell \in \mathcal{I}_{m_{\rm tot}^A,m_i^A}\\
    &\gamma_j\left(\begin{smallmatrix}d_j(\ell-m_j^G:\ell) \\ y_j(\ell-m_j^G:\ell)\end{smallmatrix}\right) \le 0,&&\forall\ell \in \mathcal{I}_{m_j^G,m_i^A},~\forall j\in \BR(i),
    \end{align*}\normalsize
    
    \vspace{-7pt}
    \hspace{-12pt}all hold, then $\alpha_i\left(\begin{smallmatrix}d_i(0: m_i^A) \\ y_i(0:m_i^A-1)\end{smallmatrix}\right) \le 0$.
    \item[ii)] If 
    
    \vspace{-15pt}
    \small
    \begin{align*}
    ~&~\alpha_{\rm tot}\left(\begin{smallmatrix}d^{\rm ext}(0:m_{\rm tot}^A) \\ y^{\rm ext} (0:m^A_{\rm tot})\end{smallmatrix}\right) \le 0,&&\forall\ell \in \mathcal{I}_{m_{\rm tot}^A,m_{\rm tot}^G},\\ ~&~\gamma_i\left(\begin{smallmatrix}d_i(\ell-m_i^G:\ell) \\ y_i(\ell-m_i^G:\ell)\end{smallmatrix}\right) \le 0,&& \forall \ell \in \mathcal{I}_{m_i^G,m_{\rm tot}^G},~\forall i\in \V,
    \end{align*}\normalsize
    
    \vspace{-7pt}\hspace{-12pt}
    all hold, then $\gamma_{\rm tot}\left(\begin{smallmatrix}d^{\rm ext} (0:m^G_{\rm tot}) \\ y^{\rm ext} (0:m^G_{\rm tot})\end{smallmatrix}\right) \le 0$.
\end{enumerate}
\end{thm_new}

The proof is found in Appendix \ref{appendix.Vert}. Colloquially, condition i) states that the assumptions of the composition $\bigotimes_{i\in \V} \C_i$ assumes less than $\C_{\rm tot}$, and condition ii) states that the composition guarantees more than $\C_{\rm tot}$. 
The theorem allows one to verify a vertical contract $\bigotimes_{i\in \V} \C_i \preccurlyeq \C_{\rm tot}$ for a feedback-less network $\G$ by verifying $|\V|+1$ implications, each of them can be cast as an LP in the variables $d^{\rm ext}(0:m),y^{\rm ext} (0:m),\{d_j(0:m),y_j(0:m)\}_{j\in\V}$:

\vspace{-23pt}
\small
\begin{subequations} \label{eq.LP_NSystemCascade}
\begin{align}
    \varrho_i = \max ~&~ \alpha_i\left(\begin{smallmatrix}d_i(0:m_i^A) \\ y_i(0:m_i^A-1)\end{smallmatrix}\right)\\ \nonumber
    {\rm s.t.}~&~ \alpha_{\rm tot}\left(\begin{smallmatrix}d^{\rm ext}(\ell-m_{\rm tot}^A:\ell) \\ y^{\rm ext} (\ell-m_{\rm tot}^A: \ell - 1)\end{smallmatrix}\right) \le 0,&&\forall \ell \in \mathcal{I}_{m_{\rm tot}^A,m_i^A},\\ \nonumber
    ~&~\gamma_j\left(\begin{smallmatrix}d_j(\ell-m_j^G:\ell) \\ y_j(\ell-m_j^G:\ell)\end{smallmatrix}\right) \le 0,&&\forall \ell \in \mathcal{I}_{m_j^G,m_i^A},\\\nonumber~&&&\forall j\in \BR(i)\\ \nonumber
    ~&~ \eqref{eq.diDefinition}\text{ at time $\ell$ and node $j$},&&\forall \ell = \mathcal{I}_{0,m_i^A}, \\\nonumber~&&&\forall j\in \BR_+(i),\\ \nonumber
    ~&~ \eqref{eq.OutputConsistency}\text{ at time $\ell$},&&\forall \ell \in \mathcal{I}_{0,m_i^A},
\end{align}

\vspace{-32pt}
\begin{align}
    \varrho_\OO = \max ~&~ \gamma_{\rm tot}\left(\begin{smallmatrix}d^{\rm ext}(0:m^G_{\rm tot}) \\ y^{\rm ext} (0:m^G_{\rm tot})\end{smallmatrix}\right)\\ \nonumber
    {\rm s.t.}~&~\alpha_{\rm tot}\left(\begin{smallmatrix}d^{\rm ext}(0:m^A_{\rm tot}) \\ y^{\rm ext} (0:m^A_{\rm tot}-1)\end{smallmatrix}\right) \le 0,&&\forall \ell \in \mathcal{I}_{m_{\rm tot}^A,m_{\rm tot}^G},\\\nonumber
    ~&~\gamma_j\left(\begin{smallmatrix}d_j(\ell-m_j^G:\ell) \\ y_j(\ell-m_j^G:\ell)\end{smallmatrix}\right) \le 0,&&\forall \ell \in \mathcal{I}_{m_j^G,m_{\rm tot}^G},\\\nonumber~&&&\forall j\in \V\\\nonumber
    ~&~ \eqref{eq.diDefinition}\text{ at time $\ell$ and node $j$},&&\forall \ell \in \mathcal{I}_{0,m_{\rm tot}^G}, \\\nonumber~&~&&\forall j\in \V\\\nonumber
    ~&~ \eqref{eq.OutputConsistency}\text{ at time $\ell$},&&\forall \ell \in \mathcal{I}_{0,m_{\rm tot}^G}.
\end{align}
\end{subequations} \normalsize
These programmes give rise to an algorithm determining whether $\bigotimes_{i\in \V} \C_i \preccurlyeq \C_{\rm tot}$, see Algorithm \ref{alg.VerifyNSysCascade}. It is an LP-based verification method for feedback-less vertical contracts, solving a total of $|\V|+1$ LPs. They can be solved using standard optimisation software. The correctness of the algorithm is stated in the following corollary:

\begin{algorithm} [t]
\caption{Verifying Vertical Contracts for Feedback-less Networks}
\label{alg.VerifyNSysCascade}
{\bf Input:} A networked system defined by a DAG $\G = (\V,\E)$, and output set $\W \subseteq \V$, component-level RD LTI contracts $\{\C_i\}_{i\in \V}$, and an RD LTI contract $\C_{\rm tot}$ on the composite system of the form \eqref{eq.NContractsCascade}.\\
{\bf Output:} A Boolean variable ${\bf b}_{\otimes,\preccurlyeq}$.
\begin{algorithmic}[1]
\State Compute $\{\varrho_i\}_{i\in \V},\varrho_\OO$ by solving the LPs \eqref{eq.LP_NSystemCascade}.
\If{$\{\varrho_i\}_{i\in \V}, \varrho_{\OO}$ are all non-positive}
\State {\bf Return} ${\bf b}_{\otimes,\preccurlyeq}$ = true.
\Else
\State {\bf Return} ${\bf b}_{\otimes,\preccurlyeq}$ = false.
\EndIf
\end{algorithmic}
\end{algorithm}

\vspace{-8pt}
\begin{cor_new} \label{thm.Opt}
Under the assumptions of Theorem \ref{thm.VertCascadeN}, Algorithm \ref{alg.VerifyNSysCascade} is always correct, i.e., $\bigotimes_{i\in \V} \C_i \preccurlyeq \C_{\rm tot}$ holds if and only if the algorithm returns ${\bf b}_{\otimes,\preccurlyeq}$ = true.
\end{cor_new}

\vspace{-15pt}
\begin{pf}
Follows from Theorem \ref{thm.VertCascadeN} and the following principle: Given functions $f,g : \X \to \R$ defined on an arbitrary space, the implication $f(x) \le 0 \implies g(x) \le 0$ holds if and only if $\max_x\{f(x) : g(x) \le 0\} \le 0$. \qed
\end{pf}

\vspace{-15pt}
\begin{exam}
We demonstrate the LP framework for a cascade of A/G contracts, for which the assumptions do not depend on the output variables. The network is given by $\G = (\V,\E)$, $\V = \{1,2\}$ and $\E = \{1\to 2\}$, where node $1$ corresponds to an open-loop controller and node $2$ corresponds to the system to be controlled. Thus, $\BR(1) = \emptyset$ and $\BR(2) = \{1\}$. Moreover, $\W = \{2\}$, so $d^{\rm ext} = d_1$, $d_2 = y_1$ and $y^{\rm ext}  = y_2$.
We verify that $\C_1 \otimes \C_2 \preccurlyeq \C_{\rm tot}$ by checking three implications:

\vspace{-10pt}
\begin{itemize}
    \item The assumptions of $\C_{\rm tot}$ imply the assumptions of $\C_1$. This is equivalent to $\varrho_1 \le 0$, where $\varrho_1$ is equal to
    
    \vspace{-15pt}
    \small
    \begin{align*}
    \max_{d_1} ~&~ \alpha_1\left(d_1(0:m_1^A)\right)\\ \nonumber
    {\rm s.t.}~&~ \alpha_{\rm tot}\left(d_1(\ell-m_{\rm tot}^A:\ell)\right) \le 0,&\forall \ell \in \mathcal{I}_{m_{\rm tot}^A,m_1^A}.
    \end{align*}\normalsize
    \item The assumptions of $\C_{\rm tot}$, plus the guarantees of $\C_1$, imply the assumptions of $\C_2$. This is equivalent to $\varrho_2 \le 0$, where $\varrho_2$ is equal to
    
    \vspace{-15pt}
    \small
    \begin{align*}
    \max_{d_i} ~&~ \alpha_2\left(d_2(0:m_2^A)\right)\\ \nonumber
    {\rm s.t.}~&~ \alpha_{\rm tot}\left(d_1(\ell-m_{\rm tot}^A:\ell)\right) \le 0,&& \forall\ell \in \mathcal{I}_{m_{\rm tot}^A,m_2^A}\\
    ~&~\gamma_1\left(\begin{smallmatrix} d_1(\ell-m_1^G:\ell)\\ d_2(\ell-m_1^G:\ell)\end{smallmatrix}\right) \le 0,&&\forall \ell \in \mathcal{I}_{m_1^G,m_2^A}.
    \end{align*}\normalsize
    \item The assumption of $\C_{\rm tot}$, plus guarantees of $\C_1$ and $\C_2$, imply the guarantees of $\C_{\rm tot}$. This is equivalent to $\varrho_{\rm tot} \le 0$, where $\varrho_{\rm tot}$ is equal to
        
    \vspace{-15pt}
    \small
    \begin{align*}
    \max_{d_i,y_2}~& \gamma_{\rm tot}\left(\begin{smallmatrix} d_1(0:m_{\rm tot}^G)\\ y_2(0:m_{\rm tot}^G)\end{smallmatrix}\right)\\ \nonumber
    {\rm s.t.}~& \alpha_{\rm tot}\left(d_1(\ell-m_{\rm tot}^A:\ell)\right) \le 0,&&\forall\ell \in \mathcal{I}_{m_{\rm tot}^A,m_{\rm tot}^A}\\
    &\gamma_1\left(\begin{smallmatrix} d_1(\ell-m_1^G:\ell)\\ d_2(\ell-m_1^G:\ell)\end{smallmatrix}\right) \le 0,&&\forall \ell \in \mathcal{I}_{m_1^G,m_{\rm tot}^A}\\
    &\gamma_2\left(\begin{smallmatrix} d_2(\ell-m_2^G:\ell)\\ y_2(\ell-m_2^G:\ell)\end{smallmatrix}\right) \le 0,&&\forall \ell \in \mathcal{I}_{m_2^G,m_{\rm tot}^A}.
    \end{align*} \normalsize
\end{itemize}
Indeed, the first and second implications above are the implication i) in Theorem \ref{thm.VertCascadeN} for the vertices $1,2\in\V$ respectively, and the third implication above is the implication ii) from Theorem \ref{thm.VertCascadeN}.
\end{exam}

\begin{rem_new} \label{eq.MultDepth}
The LP problems above depend on the depths of the RD LTI contracts. One could consider a contract with multiple assumptions or guarantees defined by different depths. In that case, the problems \eqref{eq.LP_NSystemCascade} should be amended as follows: Whenever we use the contract for defining constraints, we add different constraints for each assumption or guarantee, having different relevant times $\ell$. Whenever we use the contract for defining the cost function, replace it with the maximum of all corresponding piecewise-linear functions.
\end{rem_new}

\vspace{-13pt}
\section{Networks with Feedback} \label{sec.GeneralFeedback}
\vspace{-12pt}
The previous section focused on feedback-less networks. In this section, we generalise our results to general networks with feedback, e.g., the connection of a feedback controller to a system.

\vspace{-10pt}
\subsection{Causality and Algebraic Loops}
\vspace{-10pt}
Before delving into the definition of $\bigotimes_{i\in \V} \C_i$, we must understand its basic limitations. We demonstrate them in an example.

\vspace{-6pt}
\begin{exam} \label{exam.AlgebraicLoop}
Consider the network in Fig. \ref{fig.FeedbackTwoSystems}. with  $\C_1 = (\D_1,\OO_1)$ and $\C_2 = (\D_2,\OO_2)$ being the following RD contracts:

\begin{figure}[t]
    \centering
    \includegraphics[width = 0.27\textwidth]{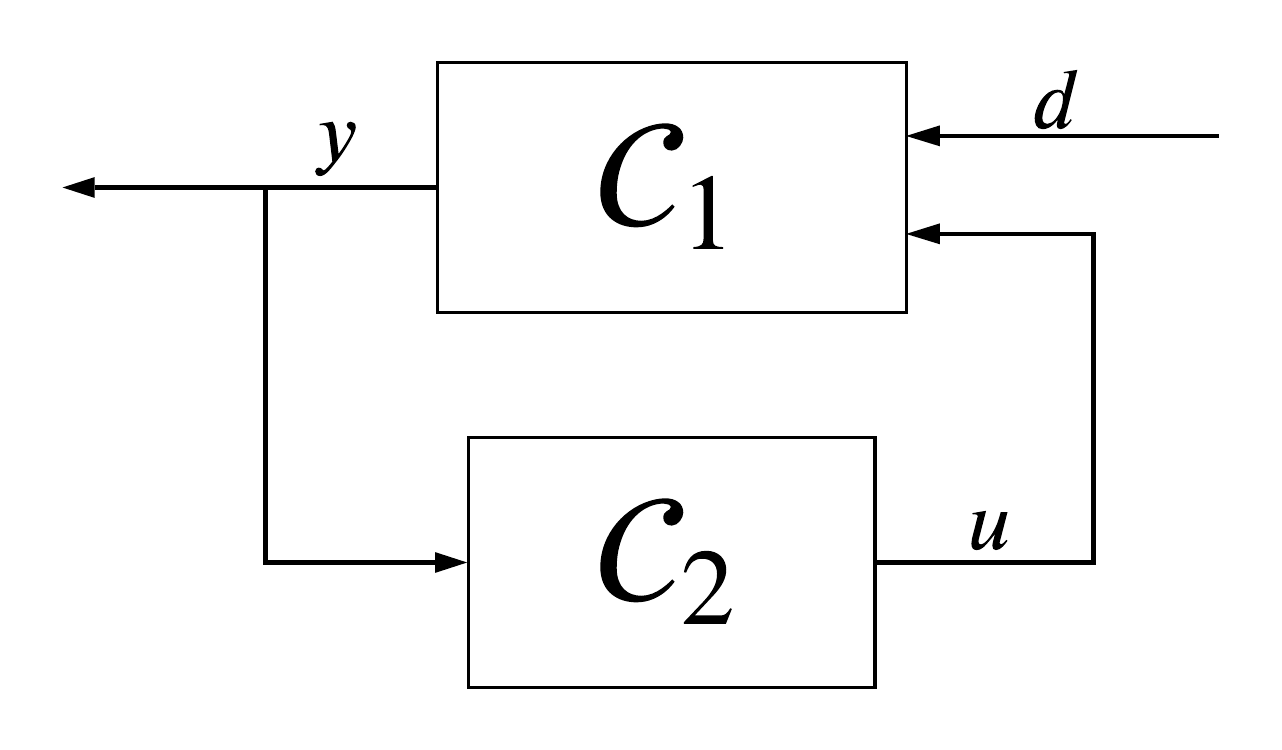}
    \vspace{-10pt}
    \caption{Feedback Composition of two contracts.}
    \label{fig.FeedbackTwoSystems}
\end{figure}

\vspace{-30pt}
\begin{align*}
    \D_1 &= \{(d(\cdot),u(\cdot)): |d(k)|\le 1~|u(k)| \le 1, \forall k\},\\
    \OO_1 &= \{(d(\cdot),u(\cdot),y(\cdot)): y(k) = (d(k) + u(k)) + 1, \forall k\},\\
    \D_2 &= \{y(\cdot): |y(k)| \le 1, \forall k\},\\
    \OO_2 &= \{(y(\cdot),u(\cdot)): u(k) = y(k) + 1, \forall k\}.
\end{align*}
If the composition $\C_1\otimes \C_2$ could be defined, and $(d,y)$ is an input-output pair satisfying its guarantees, we should have $(d,u,y)\in \OO_1, (y,u)\in \OO_2$ for some signal $u$, i.e., for any $k\in \N$, we would have $y(k) = d(k)+u(k)+1$ and $u(k) = y(k)+1$. The only solution to these equations is the constant signal $d(k) = -2$, which is not compatible with $\D_1$. Hence, $\C_1\otimes\C_2$ cannot be defined meaningfully in this case.
\end{exam}

The inconsistency in Example \ref{exam.AlgebraicLoop} arises from contradicting specifications. More precisely, the guarantees of $\C_1$ constrain $y(k)$ in terms of $u(k)$, and the guarantees of $\C_2$ constrain $u(k)$ in terms of $y(k)$, resulting in an algebraic loop creating ill-posed constraints. This situation can be avoided if we demand that $\C_1$ would constrain $y(k)$ using only $d(0:k), y(0:k-1),u(0:k-1)$ and not using $u(k)$, which can be understood as a strict causality-type demand on the contract $\C_1$ with respect to the control input $u$. This motivates the following definition:

\begin{defn_new} \label{def.RDCon}
Let $\C = (\D,\OO)$ be an RD contract of the form \eqref{eq.GCAG_A},\eqref{eq.GCAG_G} with input $d\in \Sig^{n_d}$ and output $y\in \Sig^{n_y}$. 
Suppose $d_{\rm sub}$ is a subsignal of $d$.
$\C$ is \emph{strictly} recursively defined with respect to $d_{\rm sub}$, denoted SRD($d_{\rm sub}$), if for any time $k$, the condition defining $\C$'s guarantees at time $k$, $\left[\begin{smallmatrix}d(k) \\ y(k)\end{smallmatrix}\right] \in G_k\left(\begin{smallmatrix} d(0:k-1)\\y(0:k-1)\end{smallmatrix}\right)$, is independent of  $d_{\rm sub}(k)$.\footnote{If $d_{\rm sub}^\prime$ is the complementary subsignal to $d_{\rm sub}$, the condition is equivalent to the existence of set-valued functions $\tilde{G}_k$ such that $\left[\begin{smallmatrix}d(k) \\ y(k)\end{smallmatrix}\right] \in G_k\left(\begin{smallmatrix} d(0:k-1)\\y(0:k-1)\end{smallmatrix}\right)$ holds if and only if $\left[\begin{smallmatrix}d_{\rm sub}^{\prime}(k) \\ y(k)\end{smallmatrix}\right] \in \tilde G_k\left(\begin{smallmatrix} d(0:k-1)\\y(0:k-1)\end{smallmatrix}\right)$.}
\end{defn_new}

\begin{figure}[b]
    \centering
    \includegraphics[width = 0.49\textwidth]{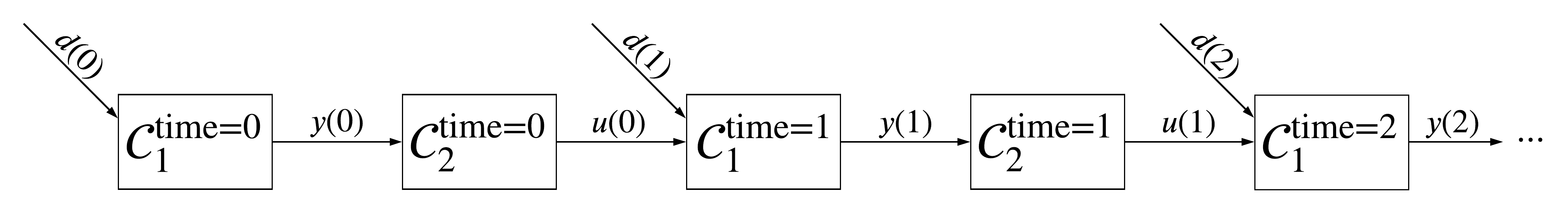}
    \caption{An infinite cascade composition, equivalent to the feedback connection in Fig. \ref{fig.FeedbackTwoSystems} if $\C_1$ is RD and $\C_2$ is SRD($u$).}
    \label{fig.InfiniteCascadeTwoSystems}
\end{figure}

As explained above, the ill-posedness issue in Example \ref{exam.AlgebraicLoop} could not occur if $\C_1$ was SRD($u$) and $\C_2$ was RD. Indeed, there is a clear ``order of constraining" guaranteeing well-posedness: in the sequence $y(0),u(0),y(1),u(1),\ldots$, each element is constrained using the preceding elements, but not using the following elements. This ``order of constraining" is illustrated in Fig. \ref{fig.InfiniteCascadeTwoSystems}, replacing the feedback composition by an infinite cascade composition. This approach can be generalised to more intricate networks. Suppose there exists an ``order of constraining" given by $y_{i_1}(0),\ldots,y_{i_N}(0)$,$y_{i_1}(1),\ldots,y_{i_N}(1),y_{i_1}(2),\ldots$. Then $y_{i_1}(k)$ is constrained by $\{y_{i_q}(0:k-1)\}_{q=1}^N$, so $\C_{i_1}$ must be SRD with respect to $y_{i_2},\ldots,y_{i_N}$. Similarly, $y_{i_2}(k)$ is only constrained by $\{y_{i_q}(0:k-1)\}_{q=1}^N$ and $y_{i_1}(k)$, implying that $\C_2$ is SRD with respect to $y_{i_3},\ldots,y_{i_N}$. 

\vspace{-5pt}
In Section \ref{subsec.CompFeedback}, we will define the contract composition $\bigotimes_{i\in \V} \C_i$ for RD contracts while assuming an ``order of constraining" exists. The remainder of this section is devoted to better understanding what is ``order of constraining". We start by translating strict causality to the language of graph theory:

\vspace{-5pt}
\begin{defn_new} \label{eq.nscEdges}
Given a graph $\G = (\V,\E)$ and component-level RD contracts $\{\C_i\}_{i\in \V}$, we say an edge $e = i\to j\in \E$ is \emph{strictly causal} if $\C_j$ is SRD($y_i$). We let $\E_{\rm sc}$ be the set of strictly causal edges, and $\E_{\rm nsc} = \E \setminus \E_{\rm sc}$ be the set of non-strictly causal edges.
\end{defn_new}

\vspace{-5pt}
In other words, the edge $i\to j$ is non-strictly causal if the guarantee on $y_j(k)$ can depend on $y_i(k)$. Mimicking the argument for feedback-less networks, $y_i(k)$ is constrained by $y_j(0:k-1)$ if $j$ is backward-reachable from $i$ in $\G$, i.e., if $j\in\BR(i)$. Similarly, $y_i(k)$ is constrained by $y_j(k)$ if $j$ is backward reachable from $i$ while only using non-strictly causal edges (i.e., in $\G_{\rm nsc}$). For convenience, we denote the backward-reachable set from $i$ in $\G_{\rm nsc}$ as $\BR_{\rm nsc}(i)$. In particular, (the lack of) contract-theoretic algebraic loops corresponds to (the lack of) cycles in $\G_{\rm nsc}$, leading to the following assumption:

\vspace{-10pt}
\begin{assump} \label{assump.NoAlgLoop}
Any cycle in the graph $\G$ contains at least one strictly causal edge, i.e., $\G_{\rm nsc}$ is a DAG.
\end{assump}

\vspace{-9pt}
\subsection{Composition} \label{subsec.CompFeedback}
\vspace{-9pt}
From now on, we fix a graph $\G = (\V,\E)$ with $N$ nodes, a set of output nodes $\W \subseteq \V$, and component-level RD contracts $\{\C_i\}_{i\in \V}$ satisfying Assumptions \ref{assump.1} and \ref{assump.NoAlgLoop}. For each $i\in \V$, we write the contract $\C_i = (\D_i, \OO_i)$ as:

\vspace{-23pt}
\begin{align} \label{eq.RDSCContractsGeneral}
    &\D_i = \left\{\left(\begin{smallmatrix} d_i(\cdot) \\ y_i(\cdot) \end{smallmatrix}\right): d_i(k) \in A_{k,i}\left(\begin{smallmatrix} d_i(0:k-1) \\ y_i(0:k-1) \end{smallmatrix}\right),\forall k\right\},\\ \nonumber
    &\OO_i = \left\{\left(\begin{smallmatrix} d_i(\cdot) \\ y_i(\cdot)\end{smallmatrix}\right): \left[\begin{smallmatrix} d_i(k) \\ y_i(k) \end{smallmatrix}\right] \in G_{k,i}\left(\begin{smallmatrix} d_i(0:k-1) \\ y_i(0:k-1) \end{smallmatrix}\right),\forall k\right\} 
\end{align}

\vspace{-15pt}
for set-valued maps $A_{k,i},G_{k,i}$, where the interconnection is defined by \eqref{eq.diDefinition} and \eqref{eq.OutputConsistency}. Drawing inspiration from the infinite cascade composition seen in Fig. \ref{fig.InfiniteCascadeTwoSystems} and postulates A) and B), we define the composition $\bigotimes_{i\in \V} \C_i$ as follows:

\vspace{-5pt}
\begin{defn_new} \label{def.FeedbackComp}
Consider a networked system with a graph $\G = (\V,\E)$, an output set $\W \subseteq \V$, and component-level RD contracts $\C_i = (\D_i,\OO_i)$ as in \eqref{eq.RDSCContractsGeneral}. Suppose Assumptions \ref{assump.1} and \ref{assump.NoAlgLoop} both hold.
The composition $\bigotimes_{i\in \V} \C_i = (\D_\otimes,\OO_\otimes)$ is a contract with input $d^{\rm ext}(\cdot)$ and output $y^{\rm ext} (\cdot)$, given by the following sets $\D_\otimes,\OO_\otimes$:

\vspace{-5pt}
\begin{itemize}
\item $(d^{\rm ext},y^{\rm ext} )\in \D_\otimes$ if for any signals $\{d_j, y_j\}_{j\in \V}$ satisfying the consistency constraints \eqref{eq.diDefinition} and \eqref{eq.OutputConsistency}, the following implication holds for any time $k\in \N$ and any $i\in \V$: If

\vspace{-15pt}
\small
    \begin{align*}
        ~&~\left[\begin{smallmatrix} d_j(\ell) \\ y_j(\ell) \end{smallmatrix}\right] \in G_{\ell,j}\left(\begin{smallmatrix} d_j(0:\ell-1) \\ y_j(0:\ell-1) \end{smallmatrix}\right),&&\forall\ell \in \mathcal{I}_{0,k},\\
        ~&&&\forall j\in \BR_{\rm nsc}(i)\\
        ~&~\left[\begin{smallmatrix} d_j(\ell) \\ y_j(\ell) \end{smallmatrix}\right] \in G_{\ell,j}\left(\begin{smallmatrix} d_j(0:\ell-1) \\ y_j(0:\ell-1) \end{smallmatrix}\right),&&\forall \ell \in \mathcal{I}_{0,k-1},\\
        ~&&&\forall j\in \BR_+(i) \setminus \BR_{\rm nsc}(i)
    \end{align*} \normalsize

\vspace{-7pt}
all hold, then $d_i(k) \in A_{k,i}\left(\begin{smallmatrix} d_i(0:k-1) \\ y_i(0:k-1) \end{smallmatrix}\right)$.
\item $(d^{\rm ext},y^{\rm ext} )\in \OO_\otimes$ if there exist signals $\{d_j,y_j\}_{j\in \V}$ such that the consistency constraints \eqref{eq.diDefinition} and \eqref{eq.OutputConsistency}, and $(d_j,y_j)\in \OO_j$ holds for all $j \in \V$.
\end{itemize}
\end{defn_new}

\vspace{-5pt}
Essentially, Definition \ref{def.FeedbackComp} mimics Definition \ref{defn.Feedbackless} by replacing the networked system with feedback with an infinite feedback-less networked system. This is done by replacing the contracts $\C_i$, with constraints defined over the entire time horizons, by "timewise" contracts $\C_i^{{\rm time}=k}$ constraining signals at time $k$. The counterpart to Theorem \ref{thm.CompFeedbackless} holds in the feedback case.

\vspace{-7pt}
\begin{thm_new} \label{thm.FeedbackComposition}
Let $\G = (\V,\E)$ be a network with output set $\W \subseteq \V$ and component-level RD contracts $\C_i$ as in \eqref{eq.RDSCContractsGeneral}, satisfying Assumptions \ref{assump.1} and \ref{assump.NoAlgLoop}. If $\{\Pi_i\}_{i\in \V}$ are causal systems such that $\Pi_i \sat\C_i$ holds for any $i\in \V$, then $\bigotimes_{i\in \V} \Pi_i \sat \bigotimes_{i\in \V} \C_i$.
\end{thm_new}

\vspace{-5pt}
We first state and prove the following lemma, linking the timewise contracts $\C_i^{{\rm time}=k}$ and $\C_i$:

\vspace{-5pt}
\begin{lem_new} \label{lem.CausalPartial}
Let $\C = (\D,\OO)$ be an RD contract of the form \eqref{eq.GCAG_A},\eqref{eq.GCAG_G}, where $\D$ is extendable, and let $\Pi$ be a causal system satisfying $\C$. Let $\hat d(\cdot) \in \Sig^{n_d}$ be any input signal in $\Sig^{n_d}$, and let $\hat y \in \Pi(\hat{d})$. If $\hat d(k) \in A_k\left(\begin{smallmatrix}\hat d(0:k-1)\\\hat y(0:k-1)\end{smallmatrix}\right)$ holds for $k\in \mathcal{I}_{0,n}$, then $\left[\begin{smallmatrix}\hat d(n)\\\hat y(n)\end{smallmatrix}\right] \in G_n\left(\begin{smallmatrix}\hat d(0:n-1)\\\hat y(0:n-1)\end{smallmatrix}\right)$.
\end{lem_new}

\vspace{-5pt}
In other words, satisfying the RD contract $\C$ is equivalent to satisfying all timewise contracts $\C^{{\rm time}=k}$.

\vspace{-15pt}
\begin{pf}
We will construct signals $d,y$ such that $d(0:n)=\hat d(0:n)$ and $y(0:n) = \hat y(0:n)$, $(d,y)\in \D$, and $y \in\Pi(d)$. We thus conclude from $\Pi \sat \C$ that $(d,y)\in \OO$, which yields the result by writing the guarantees at time $n$ and using $d(0:n)=\hat d(0:n)$ and $y(0:n) = \hat y(0:n)$.

\vspace{-5pt}
We now construct $d$ and $y$. Following Remark \ref{rem.timewise}, we denote the timewise set-valued maps $d(0:k) \mapsto y(k)$ as $\Pi_k$. We define $d(k)$ and $y(k)$ by induction on $k$. We first define $d(0:n) = \hat d(0:n)$ and $y(0:n) = \hat y(0:n)$, so that both $y(k) \in \Pi_k(d(0:k))$ and $d(k) \in A_k\left(\begin{smallmatrix}d(0:k-1)\\ y(0:k-1)\end{smallmatrix} \right)$ hold for $k\in \mathcal{I}_{0,n}$.
Now, assume $d(0:k),y(0:k)$ have been defined so that both $y(j) \in \Pi_j(d(0:j))$ and $d(j) \in A_j\left(\begin{smallmatrix}d(0:j-1)\\ y(0:j-1)\end{smallmatrix} \right)$ hold for $j\in \mathcal{I}_{0,k}$. By extendibility, the set $A_{k+1}\left(\begin{smallmatrix}d(0:k)\\ y(0:k)\end{smallmatrix} \right)$ is non-empty, and we choose $d(k+1)$ as one of its elements, as well as some $y(k+1) \in \Pi_{k+1}(d(0:k+1))$. By construction, we have $(d,y) \in \D$, $y\in \Pi(d)$,  $d(0:n) = \hat d(0:n)$ and $y(0:n) = \hat y(0:n)$. \qedwhite
\end{pf}
\vspace{-15pt}

Given Lemma \ref{lem.CausalPartial}, the proof of Theorem \ref{thm.FeedbackComposition} is nearly identical to the proof of Theorem \ref{thm.CompFeedbackless}.
The only difference is that we are using the timewise contracts $\C_i^{{\rm time}=k}$ instead of the RD contracts $\C_i$, and that gap is bridged by Lemma \ref{lem.CausalPartial}.

\if(0)
We now prove Theorem \ref{thm.FeedbackComposition}:
\begin{pf}
Let $\{\Pi_i\}_{i\in \V}$ be causal systems, and assume $\Pi_i \sat \C_i$ holds for $i\in \V$. We show the composite system $\Pi = \bigotimes_{i\in \V} \Pi_i$ satisfies the composite contract $\bigotimes_{i\in \V} \C_i = (\D_\otimes,\OO_\otimes)$. To do so, we take signals $d^{\rm ext}$ and $y^{\rm ext} $ such that $(d^{\rm ext},y^{\rm ext} ) \in \D_\otimes$ and $y^{\rm ext}  \in \Pi(d^{\rm ext})$, and show that $(d^{\rm ext},y^{\rm ext} ) \in \OO_\otimes$.

As $y^{\rm ext}  = \Pi(d^{\rm ext})$, there exist signals $\{d_i,y_i\}_{i\in \V}$ satisfying $y_i \in \Pi_i(d_i)$, as well as \eqref{eq.diDefinition} and \eqref{eq.OutputConsistency}. We show $(d_i,y_i) \in \OO_i$ holds for all $i\in \V$, proving that $(d^{\rm ext},y^{\rm ext} )\in \OO_\otimes$. More specifically, we prove $\left[\begin{smallmatrix} d_i(k) \\ y_i(k) \end{smallmatrix}\right] \in G_{k,i}\left(\begin{smallmatrix} d_i(0:k-1) \\ y_i(0:k-1) \end{smallmatrix}\right)$ holds for any component $i\in \V$ and any $k\in \N$. We do so using double induction, i.e., by simultaneously using induction on $k$ and on $i$. More precisely, we take an $\E_{\rm nsc}$-weakly topological ordering $\sigma : \mathcal{I}_{1,N} \to \V$ of $\G$, and use induction on the time $k$ and on $q \in \mathcal{I}_{1,N}$, where $i=\sigma(q)$.

We start with $k=0$, and $q = 1$, i.e., $i = \sigma(1)$. By Definition \ref{def.SCEdges} and Lemma \ref{lem.DAGSort}, $\BR_{\rm nsc}(\sigma(1)) = \emptyset$. Thus, the requirements of the implication in the definition of $\D_\otimes$ hold trivially, and we conclude that $d_{i}(0) \in A_{0,i}$. Thus, Lemma \ref{lem.CausalPartial} implies $\left[\begin{smallmatrix}d_i(0)\\y_i(0)\end{smallmatrix}\right] \in \G_{0,i}$ for $i=\sigma(1)$.

Now, consider $k=0$ and $i = \sigma(q)$ for $q \in \mathcal{I}_{2,N}$. By the induction hypothesis, $\left[\begin{smallmatrix}d_{\sigma(p)}(0)\\y_{\sigma(p)}(0)\end{smallmatrix}\right] \in \G_{0,\sigma(p)}$ holds for $p\in\mathcal{I}_{1,q-1}$. As $\BR_{\rm nsc}(i) \subseteq \{\sigma(1),\ldots,\sigma(q-1)\}$ holds from Lemma \ref{lem.DAGSort}, we conclude from the definition of $\D_\otimes$ that $d_i(0) \in A_{0,i}$ holds for $i=\sigma(q)$. Thus, Lemma \ref{lem.CausalPartial} implies that $\left[\begin{smallmatrix}d_i(0)\\y_i(0)\end{smallmatrix}\right] \in \G_{0,i}$ holds for $i=\sigma(q)$.

We now focus on the case $k \ge 1$, and start with $q = 1$. By the induction hypothesis, we have $\left[\begin{smallmatrix}d_{j}(\ell)\\y_{j}(\ell)\end{smallmatrix}\right] \in \G_{\ell,j}\left(\begin{smallmatrix} d_{j}(0:\ell-1)\\y_{j}(0:\ell-1)\end{smallmatrix}\right)$ for $\ell \in\mathcal{I}_{0,k-1}$ and for $j\in \V$. As we saw before, $\BR_{\rm nsc}(i) = \emptyset$ for $i=\sigma(1)$, so the definition of $\D_\otimes$ implies that $d_i(k) \in A_{k,i}\left(\begin{smallmatrix} d_i(0:k-1)\\y_i(0:k-1)\end{smallmatrix}\right)$. Juxtaposing this with the induction hypothesis, we conclude from Lemma \ref{lem.CausalPartial} that $\left[\begin{smallmatrix} d_i(k) \\ y_i(k) \end{smallmatrix}\right] \in G_{k,i}\left(\begin{smallmatrix} d_i(0:k-1) \\ y_i(0:k-1) \end{smallmatrix}\right)$.

Lastly, we shift our attention to the case $k \ge 1$, and $q \in \mathcal{I}_{2,N}$. By the induction hypothesis, we have $\left[\begin{smallmatrix}d_{j}(\ell)\\y_{j}(\ell)\end{smallmatrix}\right] \in \G_{\ell,j}\left(\begin{smallmatrix} d_{j}(0:\ell-1)\\y_{j}(0:\ell-1)\end{smallmatrix}\right)$ for all $(\ell,j) \in \mathcal{I}_{0,k-1}\times \V$, as well as for $(\ell,j) \in \{k\}\times \{\sigma(1),\ldots,\sigma(q-1)\}$.
As we saw, $\BR_{\rm nsc}(i) \subseteq \{\sigma(1),\ldots,\sigma(q-1)\}$ for $i=\sigma(q)$, so $d_i(k) \in A_{k,i}\left(\begin{smallmatrix} d_i(0:k-1)\\y_i(0:k-1)\end{smallmatrix}\right)$ follows from the definition of $\D_\otimes$. Together with the induction hypothesis, we conclude from Lemma \ref{lem.CausalPartial} that $\left[\begin{smallmatrix} d_i(k) \\ y_i(k) \end{smallmatrix}\right] \in G_{k,i}\left(\begin{smallmatrix} d_i(0:k-1) \\ y_i(0:k-1) \end{smallmatrix}\right)$.

By the principle of mathematical induction, we conclude that $\left[\begin{smallmatrix} d_i(k) \\ y_i(k) \end{smallmatrix}\right] \in G_{k,i}\left(\begin{smallmatrix} d_i(0:k-1) \\ y_i(0:k-1) \end{smallmatrix}\right)$ holds for all $i\in \V$ and all $k\in \N$. Thus $(d_i,y_i) \in \OO_i$, which together with the input-consistency constraints \eqref{eq.diDefinition} and the output-consistency constraints \eqref{eq.OutputConsistency} imply that $(d^{\rm ext},y^{\rm ext} ) \in \OO_\otimes$. \qed
\end{pf}

\fi

\vspace{-8pt}
\subsection{Vertical Contracts}
\vspace{-10pt}

We shift our attention to Problem \ref{prob.2}. As before, we build LP-based tools for verifying vertical contracts $\bigotimes_{i\in \V} \C_i \preccurlyeq \C_{\rm tot}$ for LTI RD contracts. We fix component-level LTI RD contracts $\C_i = (\D_i,\OO_i)$ for $i\in \V$ and an LTI RD contract $\C_{\rm tot} = (\D_{\rm tot},\OO_{\rm tot})$ on the integrated system, such that Assumption \ref{assump.1} holds. We let $\alpha_i,\alpha_{\rm tot},\gamma_i,\gamma_{\rm tot}$ be the corresponding piecewise-linear functions so that \eqref{eq.NContractsCascade} holds, and we denote $\bigotimes_{i\in\V} \C_i = (\D_\otimes,\OO_\otimes)$. As before, the vertical contract $\bigotimes_{i\in\V} \C_i \preccurlyeq \C_{\rm tot}$ is equivalent to the set inclusions $\D_\otimes \supseteq \D_{\rm tot}$ and $\OO_\otimes \cap \D_{\rm tot} \subseteq \OO_{\rm tot} \cap \D_{\rm tot}$. 

\vspace{-5pt}
\begin{thm_new} \label{thm.VertFeedbackN}
Consider a networked system with a graph $\G = (\V,\E)$ and output set $\W$. Let $\{\C_i\}_{i\in \V}, \C_{\rm tot}$ be LTI RD contracts as in \eqref{eq.NContractsCascade}, where Assumptions \ref{assump.1} and \ref{assump.NoAlgLoop} hold. Denote $\bigotimes_{i\in \V} \C_i = (\D_\otimes,\OO_\otimes)$. Under mild extendibility, assumptions\footnotemark[3] the following claims hold:

\vspace{-10pt}
\begin{itemize}
\item $\D_\otimes \subseteq \D_{\rm tot}$ holds if and only if the following implication holds for all $i\in \V$. For any signals $d_i,y_i,d^{\rm ext},y^{\rm ext}$, defined at times $\{0,1,\ldots,m_i^A\}$, if the consistency constraints \eqref{eq.diDefinition} and \eqref{eq.OutputConsistency} hold, and

\vspace{-13pt}
\small
\begin{align*}
        ~&~\alpha_{\rm tot}\left(\begin{smallmatrix} d^{\rm ext}(\ell-m_{\rm tot}^A:\ell) \\ y^{\rm ext}  (\ell-m_{\rm tot}^A : \ell-1)\end{smallmatrix}\right) \le 0,&&\forall \ell \in \mathcal{I}_{m_{\rm tot}^A,m_i^A}\\
        ~&~\gamma_j\left(\begin{smallmatrix} d_j(\ell-m_j^G:\ell) \\ y_j(\ell-m_j^G:\ell) \end{smallmatrix}\right) \le 0,&&\forall \ell \in \mathcal{I}_{m_j^G,m_i^A},\\~&&&\forall j\in \BR_{\rm nsc}(i),\\
        ~&~\gamma_j\left(\begin{smallmatrix} d_j(\ell-m_j^G:\ell) \\ y_j(\ell-m_j^G:\ell)\end{smallmatrix}\right) \le 0,&& \forall \ell \in \mathcal{I}_{m_j^G,m_i^A-1},\\~&&&\forall j\in \BR(i)\setminus\BR_{\rm nsc}(i),
\end{align*}\normalsize

\vspace{-10pt}
all hold, then $\alpha_i\left(\begin{smallmatrix} d_i(0:m_i^A) \\ y_i(0:m_i^A-1) \end{smallmatrix}\right) \le 0$.

\vspace{4pt}
\item $\OO_{\otimes} \cap \D_{\rm tot} \subseteq \OO_{\rm tot} \cap \D_{\rm tot}$ holds if and only if the following implication holds. For any signals $d_i,y_i d^{\rm ext},y^{\rm ext}$ defined at times $\{0,1,\ldots,m_{\rm tot}^G\}$, if the consistency constraints \eqref{eq.diDefinition} and \eqref{eq.OutputConsistency} hold, and

\vspace{-10pt}
\small
\begin{align*}
    ~&~\alpha_{\rm tot}\left(\begin{smallmatrix} d^{\rm ext}(\ell-m^A_{\rm tot}:\ell) \\ y^{\rm ext}  (\ell-m^A_{\rm tot} : \ell-1)\end{smallmatrix}\right) \le 0,&&\forall \ell \in \mathcal{I}_{m_{\rm tot}^A,m_{\rm tot}^G}\\
    ~&~\gamma_i\left(\begin{smallmatrix} d_i(\ell-m_i^G:\ell) \\ y_i(\ell-m_i^G:\ell) \end{smallmatrix}\right) \le 0,&&\forall \ell \in \mathcal{I}_{m_i^G,m_{\rm tot}^G}\\~&&&\forall i\in \V
\end{align*}\normalsize

\vspace{-10pt}
all hold, then $\gamma_{\rm tot}\left(\begin{smallmatrix} d^{\rm ext}(0:m_{\rm tot}^G) \\ y^{\rm ext} (0:m_{\rm tot}^G) \end{smallmatrix}\right) \le 0$.
\end{itemize}

\vspace{-3pt}
In particular, the vertical contract $\bigotimes_{i\in \V} \C_i \preccurlyeq \C_{\rm tot}$ holds if and only if the first implication holds for all $i\in \V$, and the second implication holds.
\end{thm_new}
The proof of Theorem \ref{thm.VertFeedbackN} is nearly identical to that of Theorem \ref{thm.VertCascadeN}.
Theorem \ref{thm.VertFeedbackN} shows the vertical contract $\bigotimes_{i\in \V} \C_i \preccurlyeq \C_{\rm tot}$ is equivalent to $|\V|+1$ implications between linear inequalities. As before, these can be cast as LPs:

\small
\vspace{-22pt}
\begin{subequations} \label{eq.LP_NSystemFeedback}
\begin{align}
    \varrho_i = \max ~&~ \alpha_i\left(\begin{smallmatrix}d_i(0:m_i^A) \\ y_i(0:m_i^A-1)\end{smallmatrix}\right)\\ \nonumber
    {\rm s.t.}~&~\alpha_{\rm tot}\left(\begin{smallmatrix} d^{\rm ext}(\ell-m_{\rm tot}^A:\ell) \\ y^{\rm ext}  (\ell-m_{\rm tot}^A : \ell-1)\end{smallmatrix}\right) \le 0, &&\forall \ell \in \mathcal{I}_{m_{\rm tot}^A,m_i^A}\\ \nonumber
    ~&~\gamma_j\left(\begin{smallmatrix} d_j(\ell-m_j^G:\ell) \\ y_j(\ell-m_j^G:\ell) \end{smallmatrix}\right) \le 0,&&\forall \ell \in \mathcal{I}_{m_j^G,m_i^A}\\~&&&\nonumber\forall j\in \BR_{\rm nsc}(i)\\ \nonumber
    ~&~\gamma_j\left(\begin{smallmatrix} d_j(\ell-m_j^G:\ell) \\ y_j(\ell-m_j^G:\ell)\end{smallmatrix}\right) \le0,&&\forall \ell \in \mathcal{I}_{m_j^G,m_i^A-1} \\~&&&\nonumber\forall j\in \scalebox{1}{$\BR(i)\setminus\BR_{\rm nsc}(i)$}\\   \nonumber
    ~&~ \eqref{eq.diDefinition}\text{ at time $\ell$ and node $j$},&&\forall \ell \in \mathcal{I}_{0,m_i}, \\\nonumber~&&&\forall j\in \BR_+(i)\\\nonumber
    ~&~ \eqref{eq.OutputConsistency}\text{ at time $\ell$},&&\forall \ell \in \mathcal{I}_{0,m_i},
\end{align}

\vspace{-30pt}
\begin{align}
    \varrho_\OO = \max ~&~ \gamma_{\rm tot}\left(\begin{smallmatrix}d^{\rm ext}(0:m_{\rm tot}^G:\ell) \\ y^{\rm ext} (0:m_{\rm tot}^G) \end{smallmatrix}\right)\\ \nonumber
    {\rm s.t.}~&~\alpha_{\rm tot}\left(\begin{smallmatrix} d^{\rm ext}(\ell-m_{\rm tot}:\ell) \\ y^{\rm ext}  (\ell-m_{\rm tot}:\ell-1)\end{smallmatrix}\right) \le 0, &&\forall \ell \in \mathcal{I}_{m_{\rm tot}^A,m_{\rm tot}^G}\\ \nonumber
    ~&~\gamma_i\left(\begin{smallmatrix} d_i(\ell-m_i^G:\ell) \\ y_i(\ell-m_i^G:\ell) \end{smallmatrix}\right) \le 0,&&\forall \ell \in \mathcal{I}_{m_i^G,m_{\rm tot}^G}\\~&~\nonumber&&\forall i\in \V\\ \nonumber
    ~&~ \eqref{eq.diDefinition}\text{ at time $\ell$ and node $j$},&&\forall \ell \in \mathcal{I}_{0,m_{\rm tot}^G}, \\\nonumber~&&&\forall j\in \V\\\nonumber
    ~&~ \eqref{eq.OutputConsistency}\text{ at time $\ell$ and node $j$},&&\forall \ell \in \mathcal{I}_{0,m_{\rm tot}^G}.
\end{align}
\end{subequations}
\normalsize

\vspace{-13pt}
They suggest an algorithm for determining whether $\bigotimes_{i\in \V} \C_i \preccurlyeq \C_{\rm tot}$ for general vertical contracts, see Algorithm \ref{alg.VerifyNSysFeedback}. As Algorithm \ref{alg.VerifyNSysCascade}, it is an LP-based verification method, solving a total of $|\V|+1$ LPs, and the algorithm is correct:

\vspace{-5pt}
\begin{thm_new}
Under the assumptions of Theorem \ref{thm.VertFeedbackN}, Algorithm \ref{alg.VerifyNSysFeedback} is always correct, i.e., $\bigotimes_{i\in \V} \C_i \preccurlyeq \C_{\rm tot}$ holds if and only if the algorithm returns ${\bf b}_{\otimes,\preccurlyeq}$ = true.
\end{thm_new}

\vspace{-15pt}
\begin{pf}
Similar to Corollary \ref{thm.Opt}. \qed
\end{pf}

\begin{algorithm} [b]
\caption{Verifying Vertical Contracts for General Networks}
\label{alg.VerifyNSysFeedback}
{\bf Input:} A networked system $\G = (\V,\E)$, an output set $\W \subseteq \V$, local RD LTI contracts $\{\C_i\}_{i\in \V}$ and an RD LTI contract $\C_{\rm tot}$ on the composite system of the form \eqref{eq.NContractsCascade} such that Assumptions \ref{assump.1} and \ref{assump.NoAlgLoop} hold.\\
{\bf Output:} A Boolean variable ${\bf b}_{\otimes,\preccurlyeq}$.
\begin{algorithmic}[1]
\State Compute $\{\varrho_i\}_{i\in \V},\varrho_\OO$ by solving the LPs \eqref{eq.LP_NSystemFeedback}.
\If{$\{\varrho_i\}_{i\in \V}, \varrho_{\OO}$ are all non-positive}
\State {\bf Return} true.
\Else
\State {\bf Return} false.
\EndIf
\end{algorithmic}
\end{algorithm}

\vspace{-15pt}
\begin{exam}
We elucidate the LP framework for general networks by demonstrating it on the feedback composition in Fig. \ref{fig.FeedbackTwoSystems}. The network is given by $\G = (\V,\E)$, $\V = \{1,2\}$, $\E_{\rm nsc} = \{1\to 2\}$, $\E_{\rm sc} = \{2\to 1\}$, $\E = \E_{\rm sc} \cup \E_{\rm nsc}$, and $\W = \{1\}$. Node $1$ corresponds to the plant and node $2$ to the feedback controller. In this case, $\BR_{\rm nsc}(1) = \emptyset$, $\BR_{\rm nsc}(2) = \{1\}$ and $\BR(1) = \BR(2) = \{1,2\}$. 
Following Fig. \ref{fig.FeedbackTwoSystems}, we denote the external input by $d$, the output of $\C_1$ as $y$, and the output of $\C_2$ by $u$. For simplicity, we consider SRD LTI contracts $\C_1,\C_2,\C_{\rm tot}$ for which the assumptions of $\C_2$ and $\C_{\rm tot}$ do not depend on the previous outputs $y$, and the guarantee of $\C_1$ depends only on $d$ and $y$. This assumption corresponds to a situation in which $\C_1$ defines an unregulated physical system, $\C_2$ defines a controller, and $\C_{\rm tot}$ defines the closed-loop system. Thus, $d_1(\cdot) = \left(\begin{smallmatrix} d(\cdot) \\ u(\cdot) \end{smallmatrix}\right)$, $d_2(\cdot) = y(\cdot)$ and $y^{\rm ext} (\cdot) = y(\cdot)$.  In order to verify that $\C_1 \otimes \C_2 \preccurlyeq \C_{\rm tot}$, we have to verify three implications:
\begin{itemize}

\vspace{-6pt}
\item If the assumptions of $\C_{\rm tot}$ hold until a certain time $n$, and the guarantees of both $\C_1,\C_2$ hold until time $n-1$, then the assumptions of $\C_1$ hold at time $n$. This is equivalent to $\varrho_1 \le 0$, where $\varrho_1$ is equal to

\vspace{-12pt}
\small
    \begin{align*}
    \max ~&~ \alpha_1\left(\begin{smallmatrix}d(0:m_1^A) \\ u(0:m_1^A-1)\end{smallmatrix}\right)\\ \nonumber
    {\rm s.t.}~&~\alpha_{\rm tot}(d(\ell-m_{\rm tot}^A:\ell)) \le 0,&&\forall \ell \in \mathcal{I}_{m_{\rm tot}^A,m_1^A}\\ \nonumber
    ~&~\gamma_1\left(\begin{smallmatrix} d(\ell-m_1^G:\ell) \\  y(\ell-m_1^G:\ell) \end{smallmatrix}\right) \le0,&& \forall \ell \in \mathcal{I}_{m_1^G,m_1^A-1}\\ \nonumber
    ~&~\gamma_2\left(\begin{smallmatrix} y(\ell-m_2^G:\ell) \\ u(\ell-m_2^G:\ell)\end{smallmatrix}\right) \le0,&& \forall \ell \in \mathcal{I}_{m_2^G,m_1^A-1}.
    \end{align*}\normalsize
    
\vspace{-3pt}
\item If the assumptions of $\C_{\rm tot}$ and guarantees of $\C_1$ hold until some time $n$, and the guarantees of $\C_2$ hold until time $n-1$, then the assumptions of $\C_2$ hold at time $n$. This is equivalent to $\varrho_2 \le 0$, where $\varrho_2$ is

\vspace{-12pt}
\small
    \begin{align*}
    \max ~&~ \alpha_2(y(0:m_2^A))\\ \nonumber
    {\rm s.t.}~&~\alpha_{\rm tot}(d(\ell-m_{\rm tot}^A:\ell)) \le 0,&&\forall \ell \in \mathcal{I}_{m_{\rm tot}^A,m_2^A}\\ \nonumber
    ~&~\gamma_1\left(\begin{smallmatrix} d(\ell-m_1^G:\ell) \\  y(\ell-m_1^G:\ell) \end{smallmatrix}\right) \le0,&& \forall \ell \in \mathcal{I}_{m_1^G,m_2^A}\\ \nonumber
    ~&~\gamma_2\left(\begin{smallmatrix} y(\ell-m_2^G:\ell) \\ u(\ell-m_2^G:\ell)\end{smallmatrix}\right) \le0,&& \forall \ell \in \mathcal{I}_{m_2^G,m_2^A-1}.
    \end{align*} \normalsize
\item The assumption of $\C_{\rm tot}$, plus guarantees of $\C_1$ and $\C_2$, imply the guarantees of $\C_{\rm tot}$. This is equivalent to $\varrho_{\rm tot} \le 0$, where $\varrho_{\rm tot}$ is given by

\vspace{-12pt}
\small
\begin{align*}
    \max ~&~ \gamma_{\rm tot}\left(\begin{smallmatrix} d(0:m_{\rm tot}^G) \\ y(0:m_{\rm tot}^G)\end{smallmatrix}\right)\\ \nonumber
    {\rm s.t.}~&~\alpha_{\rm tot}(d(\ell-m_{\rm tot}^A:\ell)) \le 0, &&\forall \ell \in \mathcal{I}_{m_{\rm tot}^A,m_{\rm tot}^G}\\ \nonumber
    ~&~\gamma_1\left(\begin{smallmatrix} d(\ell-m_1^G:\ell) \\ y(\ell-m_1^G:\ell) \end{smallmatrix}\right) \le0,&& \forall \ell \in \mathcal{I}_{m_1^G,m_{\rm tot}^G}\\ \nonumber
    ~&~\gamma_2\left(\begin{smallmatrix} y(\ell-m_2^G:\ell) \\ u(\ell-m_2^G:\ell)\end{smallmatrix}\right) \le0,&& \forall \ell \in \mathcal{I}_{m_2^G,m_{\rm tot}^G}.
    \end{align*}\normalsize
\end{itemize}
\end{exam}

As before, we can extend the framework to the case where some of the RD LTI contracts have multiple assumptions or guarantees of different depths, see Remark~\ref{eq.MultDepth}.

\vspace{-10pt}
\section{Numerical Example} \label{sec.CaseStudy}
\vspace{-10pt}

In this section, we apply the presented contract-based framework to autonomous vehicles in an $M$-vehicle platooning-like scenario. We first define the scenario and the specifications in the form of a contract. We then use the presented framework to refine the contract on the integrated $M$-vehicle system to a collection of contracts on the physical and control subsystems of each of the vehicles. Different values of $M$ will be considered to demonstrate the scalability of the approach. Lastly, we demonstrate the modularity achieved by these processes by presenting options for realising the controller subsystem satisfying the corresponding contract, and show using simulation that the specifications on the integrated system are met for the case of $M=2$ vehicles.

\vspace{-10pt}
\subsection{Scenario Description and Vertical Contracts}
\vspace{-10pt}

We consider $M$ vehicles driving along a single-lane highway. The first vehicle in the group is called the leader, and the other $M-1$ vehicles, are the followers. We are given a headway $h>0$, and a speed limit $V_{\rm max}^f$, and our goal is to verify that each of the followers keeps at least the given headway from its predecessor, and obeys the speed limit. 

We denote the position and velocity of the $i$-th vehicle in the group as $p_i$ and $v_i$ respectively.
We consider all followers as one integrated system, whose input is $d^{\rm ext} = [p_1,v_1] \in \Sig^2$ and output $y^{\rm ext}  = [p_2,v_2,\ldots,p_M,v_M] \in \Sig^{2(M-1)}$. The guarantees can be written as $p_{r-1}(k) - p_r(k) - hv_r(k) \ge 0$ and $0\le v_r(k) \le V_{\rm max}^f$ for any $k\in \N$ and $r\in \{2,\ldots,M\}$. We assume the leader follows the first kinematic law, i.e., $p_1(k+1) = p_1(k) + \Delta t v_1(k)$ holds for any time $k$, where $\Delta t>0$ is the length of the discrete time-step. We further assume the leader obeys a speed limit $V_{\rm max}^l$, i.e., that $0 \le v_1(k) \le V_{\rm max}^l$ holds for $k\in \N$. The assumptions and guarantees define a contract $\C_{\rm tot}$ on the followers. For this example, we take $\Delta t = 1[{\rm s}], h = 2[{\rm s}$], $V_{\rm max}^l = 110[{\rm km/h}]$ and $V_{\rm max}^f = 100[{\rm km/h}]$.

\begin{figure}[b]
    \centering
    \includegraphics[width = 0.45\textwidth]{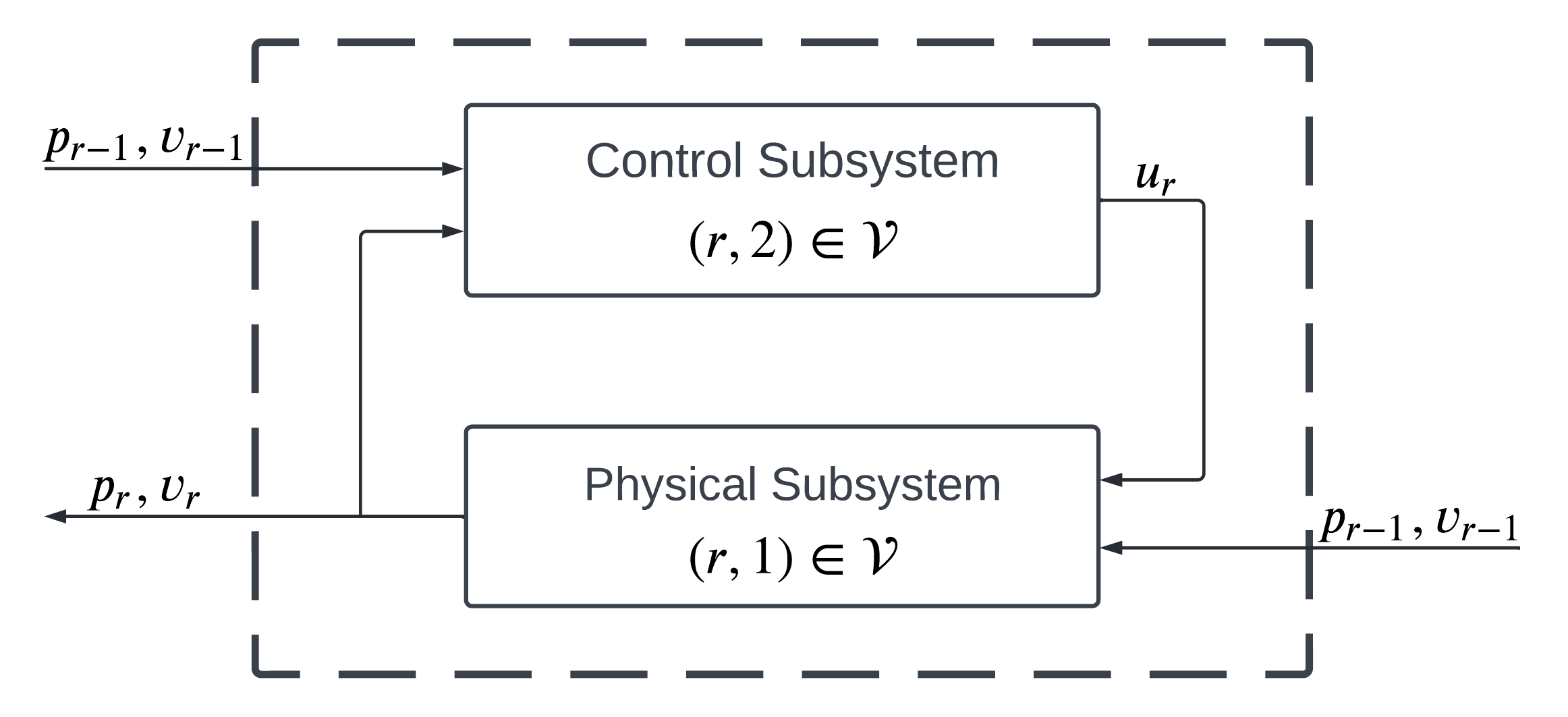}
    \vspace{-5pt}
    \caption{Interconnection topology of the $r$-th follower, for $r\ge 2$.}
    \label{fig.InterconnectionNumEx}
\end{figure}

We consider each follower vehicle as the interconnection of two subsystems in feedback:  a physical subsystem, including all physical components, actuators, etc.; and a control subsystem, which measures the physical subsystem and the environment, and issues a control signal to the physical components. The interconnection of the two systems composing the $i$-th follower can be seen in Fig. \ref{fig.InterconnectionNumEx}.
The following paragraphs describe the inputs, outputs, assumptions and guarantees associated with the ``local" contracts on each subsystem.

First, we consider the physical subsystem, corresponding to the vertex $i=(r,1)\in \V$. Intuitively, the input should only include the control input $u_r$. However, the headway guarantee refers to the position and velocity of the $(r-1)$-th vehicle. Thus, we take the input $d_i = [p_{r-1},v_{r-1},u_r]$ and the output $y_i = [p_r,v_r]$. The physical subsystem is associated with a contract $\C_{{\rm phy},r}$. We assume that the $(r-1)$-th vehicle follows the kinematic law $p_{r-1}(k+1) = p_{r-1}(k) + \Delta t v_{r-1}(k)$ for any $k\in \N$. Moreover, we assume the control input satisfies the following inequalities:

\vspace{-27pt}
\begin{align*}
        &u_r \le \frac{p_{r-1}-p_r - hv_r}{h\Delta t} + \frac{v_{r-1} - v_r}{h} - \omega_{\rm acc},\\
        &\frac{- v_r}{\Delta t} + \omega_{\rm acc}\le  u_r \le \frac{V_{\rm max}^f - v_r}{\Delta t} - \omega_{\rm acc}.
\end{align*}

\vspace{-17pt}
Here, $\omega_{\rm acc}$ is a bound on the parasitic acceleration due to wind, friction, etc., which is taken as $\omega_{\rm acc} = 0.3[{\rm m/s^2}]$. These bounds on the control input are motivated by realistic conditions, see \cite{Sharf2021b,SharfADHS2020}. As for guarantees, we desire that the headway and speed limit are kept, i.e., that $p_{r-1}(k) - p_r(k) - hv_r(k) \ge 0$ and $0 \le v_r(k) \le V_{\rm max}^f$ hold for all $k\in \N$. We also specify a guarantee that the follower satisfies $p_r(k+1) = p_r(k) + \Delta t v_r(k)$. Thus, $\C_{{\rm phy},r}$ is an SRD contract which is strictly causal with respect to $u_r$, as the guarantees at time $k$ are independent of $u_r(k)$.

Second, we consider an SRD contract $\C_{{\rm ctr},r}$ on the control subsystem, matching the vertex $i=(r,2)\in \V$. The input includes the position and velocity of both the $r$-th and the $(r-1)$-th vehicles, i.e., $d_i = [p_{r-1},v_{r-1},p_r,v_r]$, and its output is $y_i = u_r$. The contract assumes both vehicles follow the kinematic relations and the speed limits, i.e., that $p_{r-1}(k+1) = p_{r-1}(k) + \Delta t v_{r-1}(k)$, $p_r(k+1) = p_r(k) + \Delta t v_r(k)$, $v_r(k+1) = v_r(k) + \Delta t a_r(k)$, $0\le v_{r-1}(k) \le V_{\rm max}^l$ and $0\le v_r(k) \le V_{\rm max}^f$ all hold for any time $k\in \N$. These assumptions can be understood as working limitations for the sensors used by the subsystem to measure the environment, or as first principles used to generate a more exact estimate of the state, which is used for planning and the control law. For guarantees, the control signal $u_r$ must satisfy at any time $k$:

\vspace{-28pt}
\begin{align} \label{eq.ControllerContract}
        &u_r \le \frac{p_{r-1}-p_r-hv_r}{h\Delta t} + \frac{v_{r-1}-v_r}{h} - \omega_{\rm acc},\nonumber\\
        &\frac{- v_r}{\Delta t} + \omega_{\rm acc} \le u_r \le \frac{V_{\rm max}^f - v_r}{\Delta t} - \omega_{\rm acc}.
\end{align}

\begin{figure}[b]
    \centering
    \includegraphics[width = 0.48\textwidth]{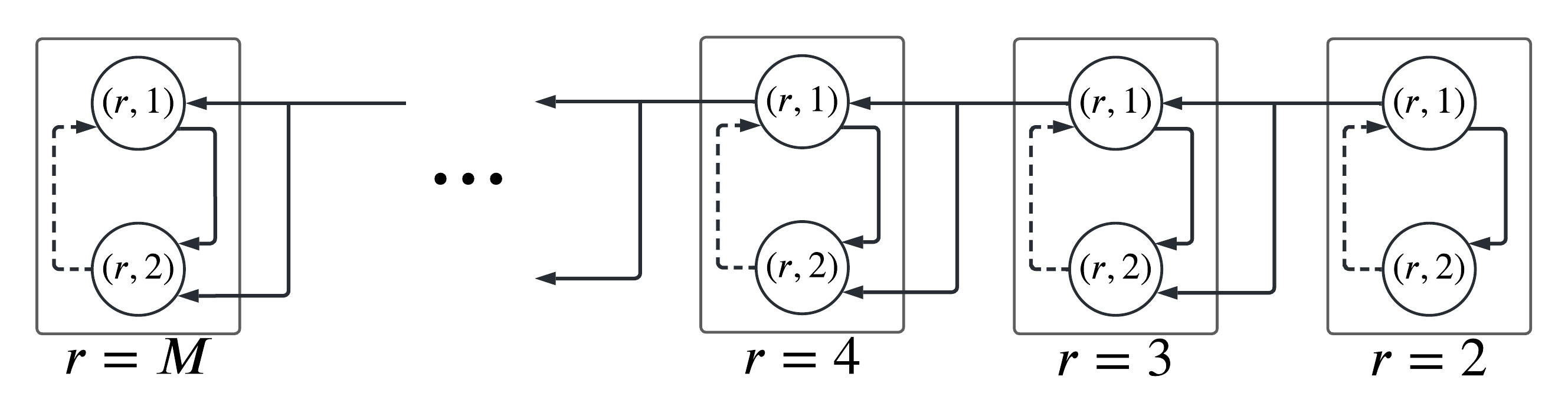}
    \vspace{-12pt}
    \caption{The interconnection graph $\G = (\V,\E)$ for the scenario in Section \ref{sec.CaseStudy} with $M$ vehicles. Dashed lines correspond to strictly causal edges, and solid lines correspond to non-strictly causal edges. Each square aggregates the subsystems corresponding vehicle $\# r$.}
    \label{fig.GraphNVehicles}
\end{figure}

\vspace{-17pt}
We wish to prove that the composition of $\C_{{\rm phy},r}$ and $\C_{{\rm ctr},r}$ for $r\in \{2,\ldots,M\}$ refines $\C_{\rm tot}$, and we do so using Algorithm \ref{alg.VerifyNSysFeedback}. First, the networked system is modelled by a graph $\G = (\V,\E)$ with $\V = \{(r,j) : r\in \{2,\ldots,M\}, j\in \{1,2\}\}$. As seen in Fig. \ref{fig.GraphNVehicles}, the set $\E_{\rm nsc}$ includes the edges $(r,1)\to(r,2)$, as well as the edges $(r-1,1)\to(r,1)$ and $(r-1,1)\to(r,2)$ for $r\in \{3,\ldots,M\}$. The set $\E_{\rm sc}$ includes the edges $(r,2)\to(r,1)$ for $r\in\{2,\ldots,M\}$. An illustration of $\G$ can be seen in Fig. \ref{fig.GraphNVehicles}, which shows the network has no algebraic loops. As the output of $\C_{\rm tot}$ includes the position and velocity of all followers, we take $\W = \{(r,1): r\in \{2,\ldots,M\}\}$. Thus, running Algorithm \ref{alg.VerifyNSysFeedback} requires us to solve a total of $|\V|+1 = 2M-1$ LPs. We solve them using MATLAB's LP solver for different values of $M$, detailed in Table \ref{table.Runtime}. In all cases, we find $\varrho_i = \varrho_\OO = 0$ for all $i\in \V$, so the vertical contract $\bigotimes_{r\in \{2,\ldots,M\}} (\C_{{\rm phy},r}\otimes \C_{{\rm ctr},r}) \preccurlyeq \C_{\rm tot}$ holds. In all cases, the algorithm was run using a Dell Latitude 7400 computer with an Intel Core i5-8365U processor, and the runtimes are reported in Table \ref{table.Runtime}. The results are further discussed below.

\begin{table*}[!th]
\begin{center}
\begin{tabular}{ |c|c|c|c|c|c|c|c| } 
\hline
$M$ & $|\V|$ & Num. of LP & Avg. Var. Num. & Avg. Constraint Num. & Network Time [s] & LP Time [s] & Total Time [s]\\ \hline
2 & 2 & 3 & 14.00 & 13.33 & 0.33 & 1.52 & 1.86\\\hline
5 & 8 & 9 & 31.33 & 21.11 & 0.35 & 1.67 & 2.02\\ \hline
10 & 18 & 19 & 56.95 & 31.58 & 0.38 & 2.15 & 2.54 \\ \hline
20 & 38 & 39 & 107.23 & 51.79 & 0.42 & 4.33 & 4.75 \\ \hline
50 & 98 & 99 & 257.39 & 111.92 & 0.57 & 31.11 & 31.69\\ \hline
100 & 198 & 199 & 507.45 & 211.96 & 0.83 & 287.76 & 288.60\\ \hline

\end{tabular}
\end{center}
\caption{An analysis of the runtime of Algorithm \ref{alg.VerifyNSysFeedback} for the vertical contract detailed in Section \ref{sec.CaseStudy}. Network Time refers to the time it took to compute the sets $\BR,\BR_{\rm nsc}$ needed to define the LPs. LP Time refers to the time it took to assemble and solve the LPs using MATLAB's own LP solver.}
\vspace{-7pt}
\label{table.Runtime}
\end{table*}

\vspace{-8pt}
\subsection{Demonstrating Modularity via Simulation}
\vspace{-8pt}
In this section, we focus on the case $M=2$, and thus drop the index $r$ from the contracts $\C_{{\rm phy},r}$ and $\C_{{\rm ctr},r}$ and from $u_r$. In this case, the vertical contract $\C_{\rm phy}\otimes \C_{\rm ctr} \preccurlyeq \C_{\rm tot}$ can be interpreted as follows: if the physical and control subsystems of the single follower are designed to satisfy $\C_{\rm phy}$ and $\C_{\rm ctr}$, then the integrated system satisfies $\C_{\rm tot}$. The two subsystem-level contracts are independent of each other, meaning these subsystems can be independently analysed, designed, verified, and tested. We demonstrate this fact by choosing a realisation for the physical subsystems, as well as two realisations for the control, and running the closed-loop system in simulation to show the guarantees hold for both control laws. 

\vspace{-6pt}
For the physical subsystem, we consider a double integrator $p_2(k+1) = p_2(k) + \Delta t v_2(k), v_2(k+1) = v_2(k) + \Delta t(u(k) + \omega_a(k))$ with acceleration uncertainty. For the realisation $\Sigma_{\rm phy}$, acceleration uncertainty is taken as i.i.d. uniformly distributed between $-\omega_{\rm acc}$ and $\omega_{\rm acc}$. It can be verified that $\Sigma_{\rm phy}\sat \C_{\rm phy}$ using $k$-induction, similarly to the framework presented in \cite{Sharf2021b}.

\vspace{-6pt}
For the control subsystem, the first realisation $\Sigma^{(1)}_{\rm ctr}$ is achieved by taking $u(k)$ as the minimum of the two upper bounds in \eqref{eq.ControllerContract}. The second realisation chooses $u(k)$ using an MPC-like controller over a horizon of $T = 5$ steps, assuming constant velocity for the leader. More precisely, $u(k) = u^0$ is chosen by optimising $\sum_{t=1}^T[(v_2^t - V_{\rm des})^2 + (u^t - u^{t-1})^2]$ over the variables $\{p_1^t,v_1^t,p_2^t,v_2^t,u^t\}_{t=0}^T$, under the input constraints \eqref{eq.ControllerContract}, the kinematic rules $p_1^{t+1} = p_1^{t} + \Delta v_1^t, v_1^{t+1} = v_1^t, p_2^{t+1} = p_2^{t} + \Delta v_2^t$ and $v_2^{t+1} = v_2^{t} + \Delta u^t$, and the initial constraints $p_2^0 = p_2(k), p_1^0 = p_1(k), v_2^0 = v_2(k)$ and $v_1^0 = v_1(k)$. For the simulation, we choose $V_{\rm des} = 90[{\rm km/h}]$. It can be verified that both systems satisfy $\C_{\rm ctr}$.

\begin{figure*}[t]
    \centering
    \subfigure[Leader velocity] {\scalebox{.40}{\includegraphics{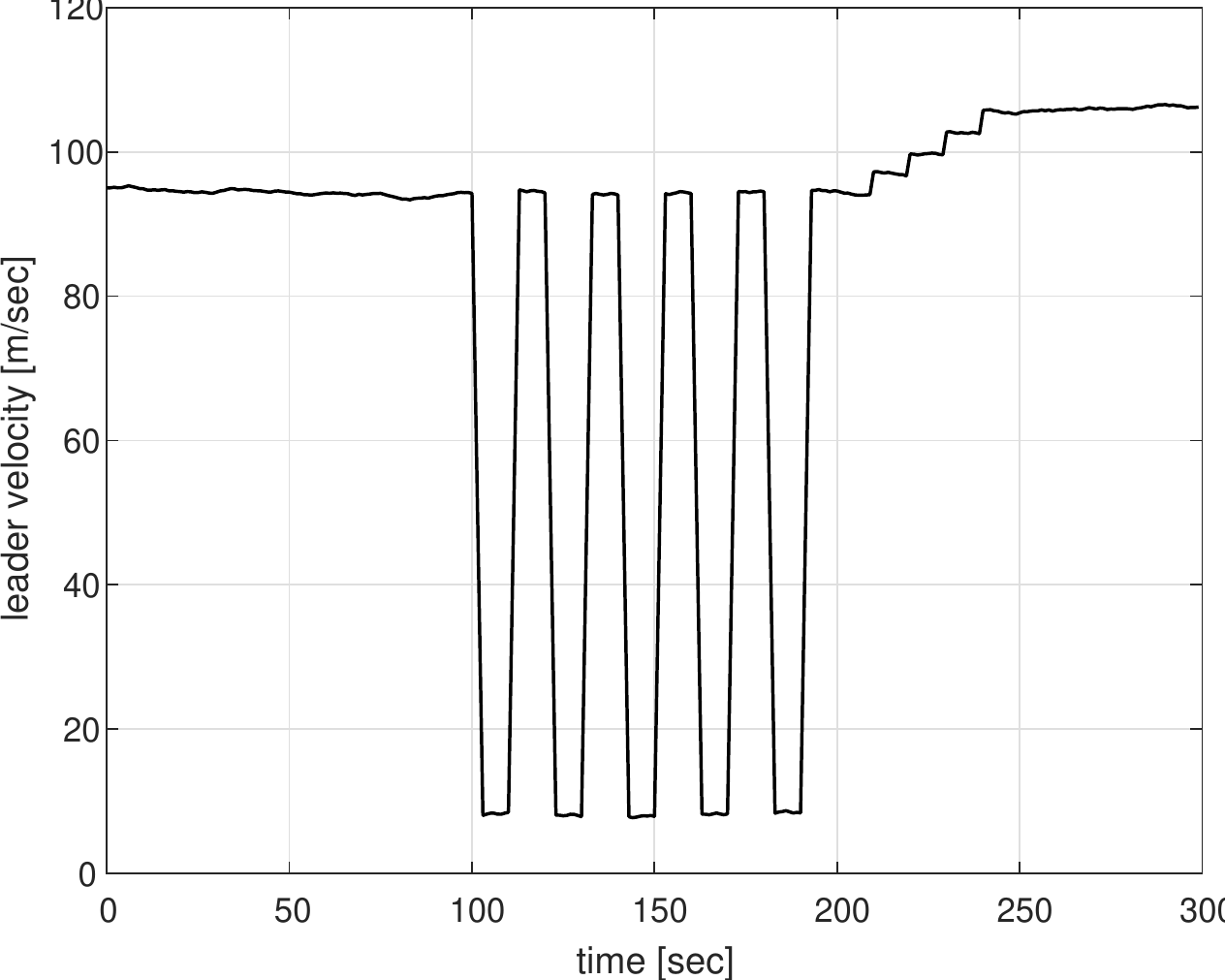}}}\hspace{.5cm}
    \subfigure[Leader acceleration] {\scalebox{.40}{\includegraphics{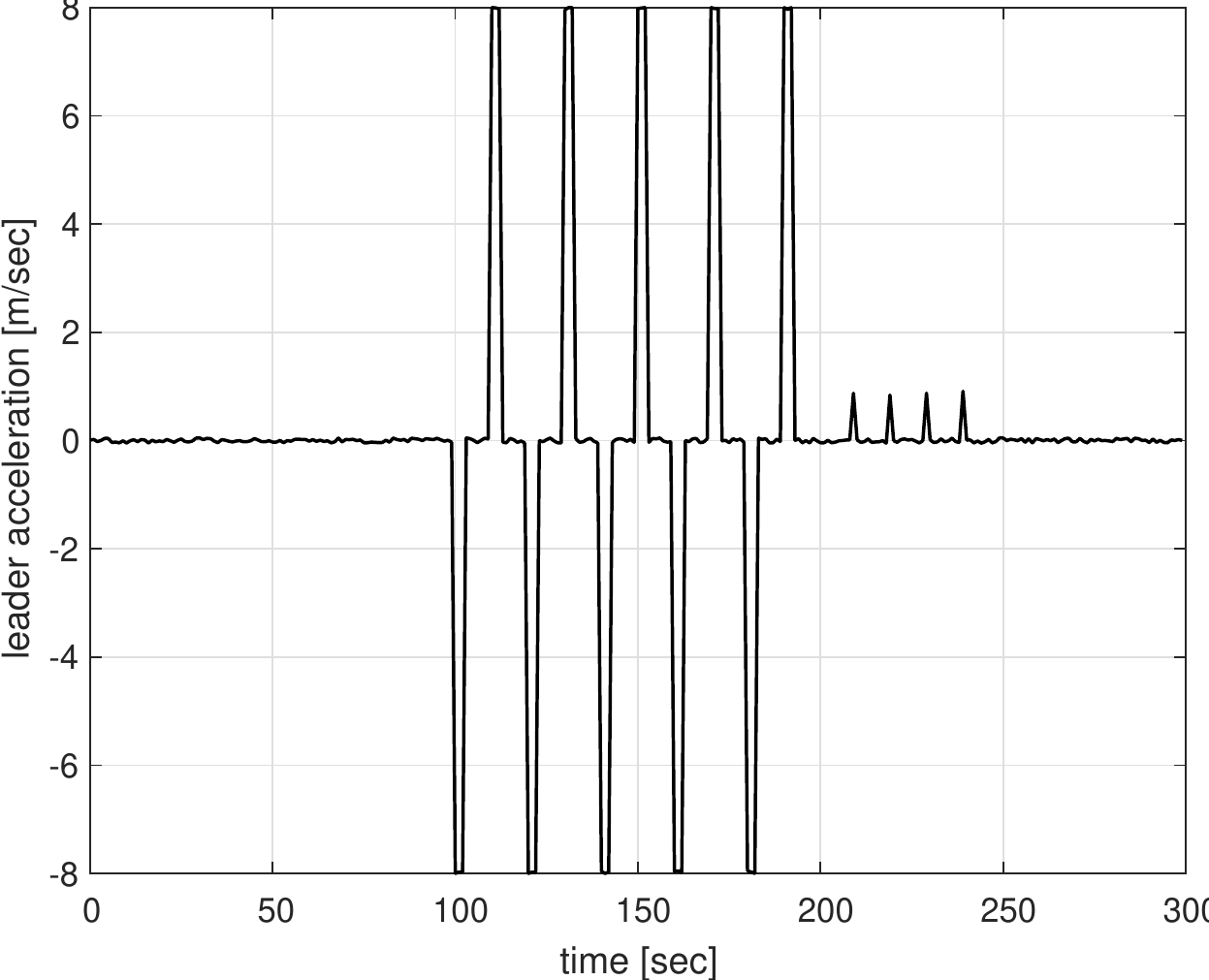}}}\hspace{.5cm}
    \subfigure[Headway, first run] {\scalebox{.40}{\includegraphics{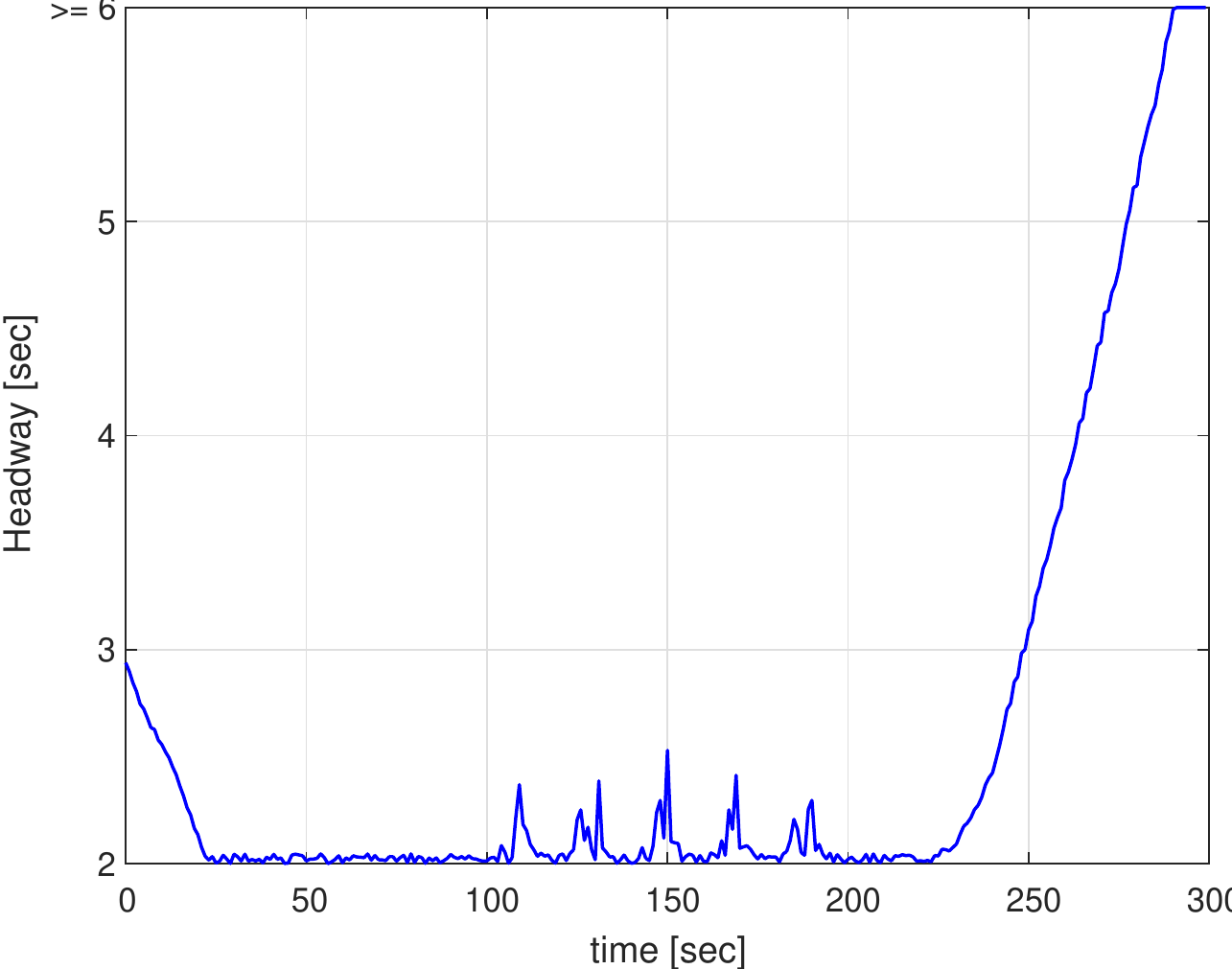}}}\hspace{2cm}
    \subfigure[Follower velocity, first run] {\scalebox{.40}{\includegraphics{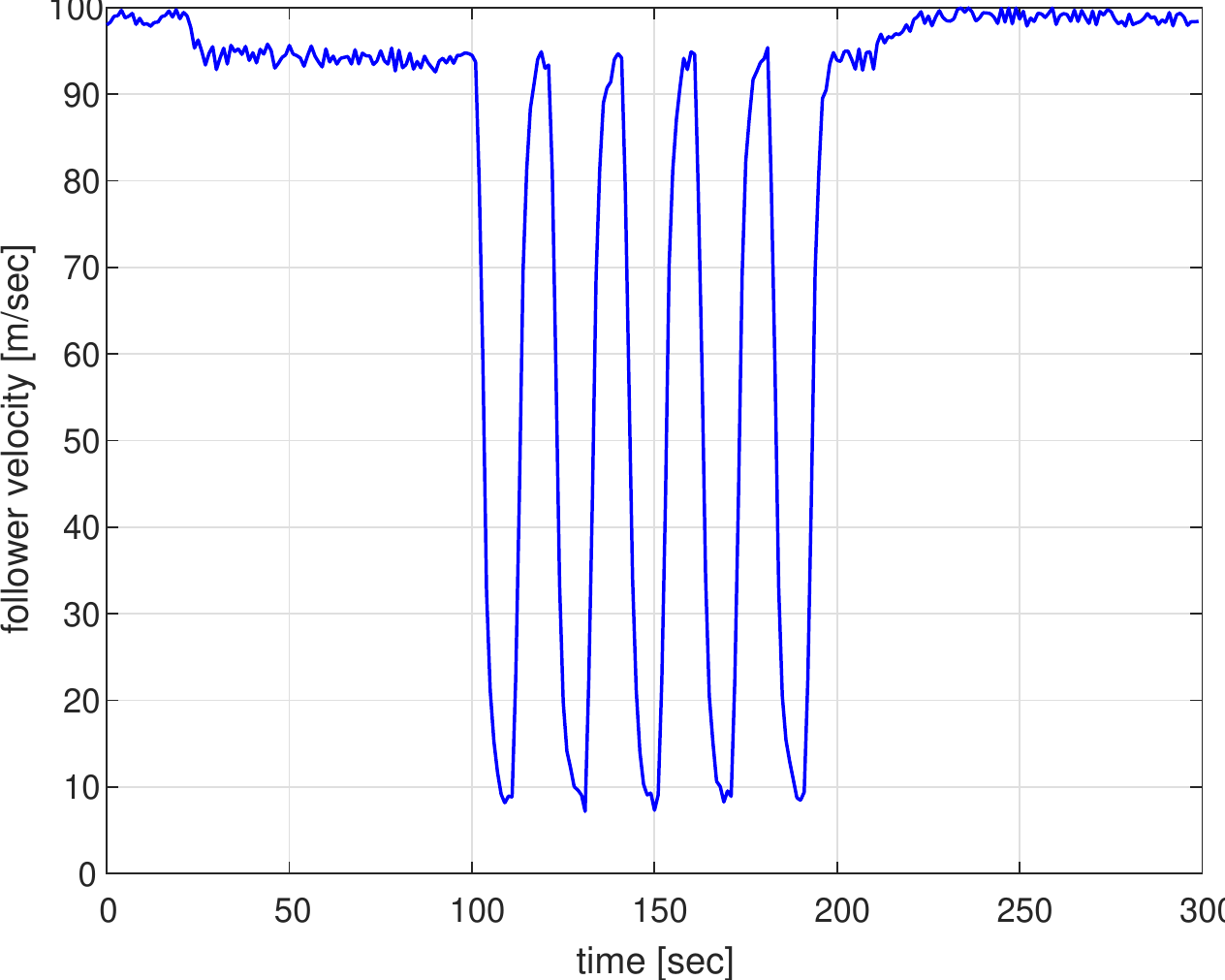}}}\hspace{.5cm}
    \subfigure[Headway, second run] {\scalebox{.40}{\includegraphics{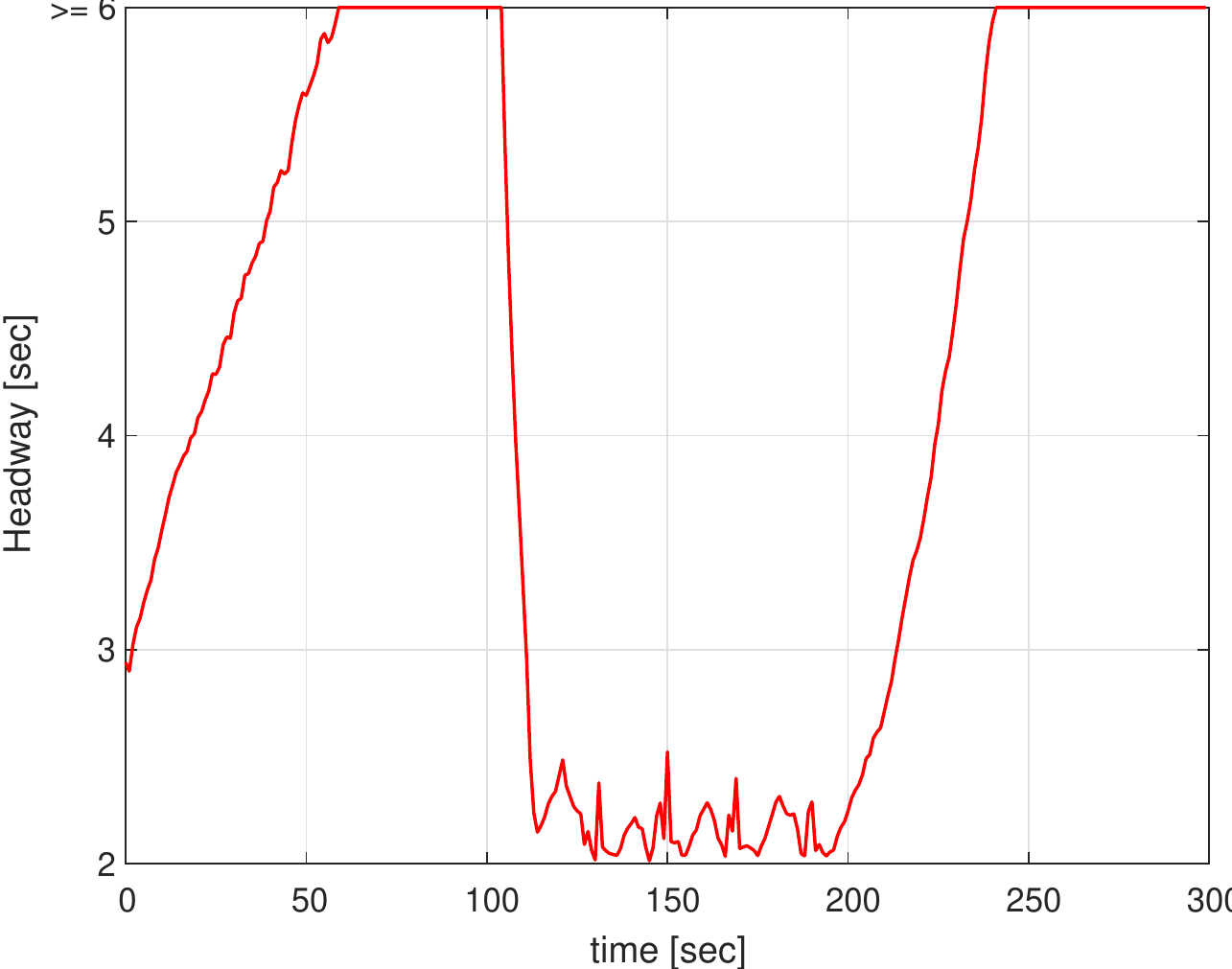}}}\hspace{.5cm}
    \subfigure[Follower velocity, second run] {\scalebox{.40}{\includegraphics{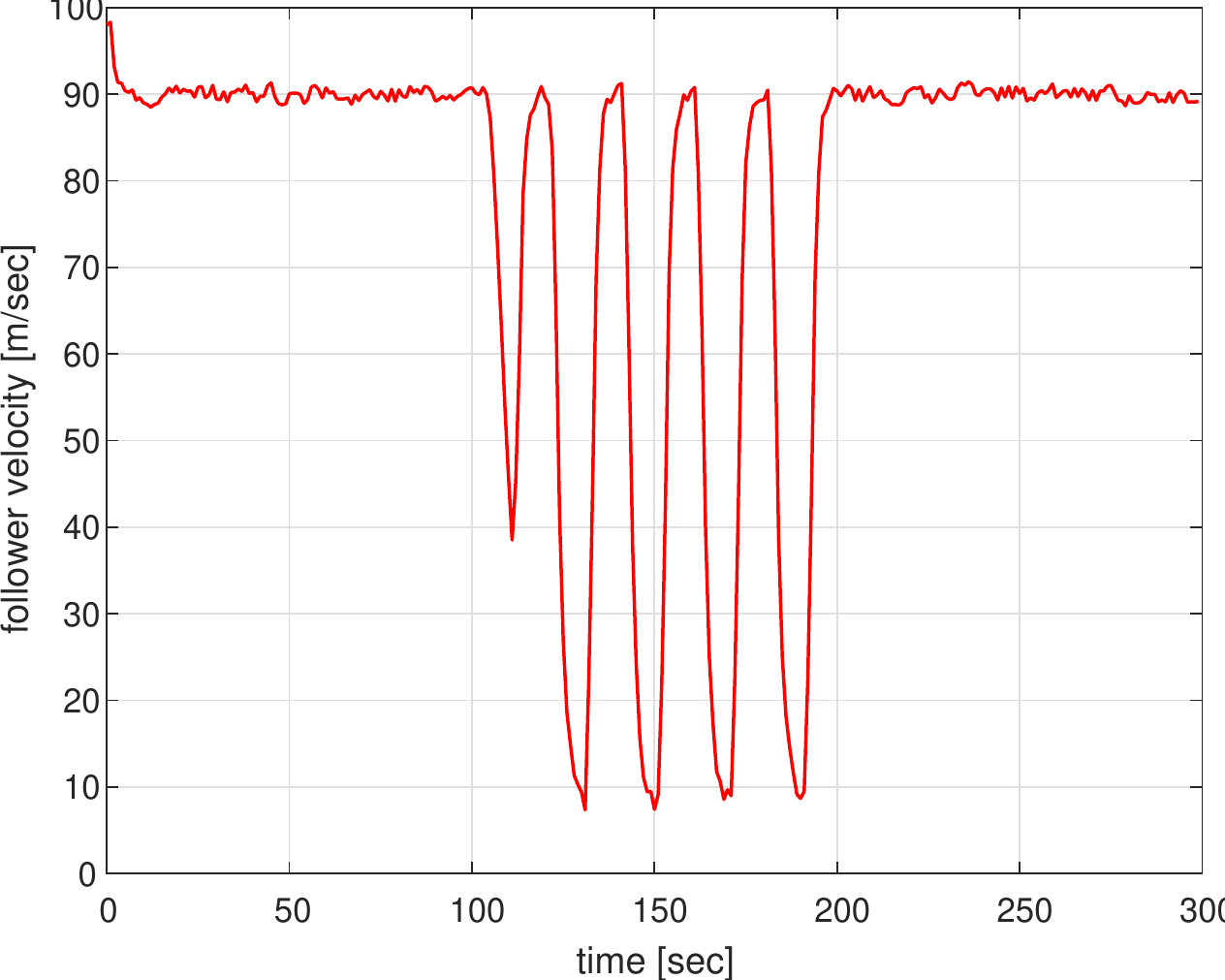}}}
    \vspace{-6pt}
    \caption{Simulation of the two-vehicle leader-follower system. The black plots correspond to the leader, the blue plots correspond to the first run of the simulation, and the red plots corresponds to the second run of the simulation.}
    \vspace{-7pt}
    \label{fig.LeaderSimulation}
\end{figure*}

\vspace{-6pt}
Both realisations $\Sigma_{\rm phy} \otimes \Sigma^{(1)}_{\rm ctr}$ and $\Sigma_{\rm phy} \otimes \Sigma^{(2)}_{\rm ctr}$ satisfy $\C_{\rm tot}$. We run simulations of length $300[{\rm s}]$. In the simulations, the leader starts at a speed of $95[{\rm km/h}]$, and $80[{\rm m}]$ in front of the follower, having an initial speed of $98[{\rm km/h}]$. The leader will roughly keep its speed for the first $100$ seconds. In the next $100$ seconds, it will brake and accelerate hard, repeatedly changing its velocity between $95[{\rm km/h}]$ and $10[{\rm km/h}]$. For the last $100$ seconds of the simulations, the leader slowly accelerates to about $105[{\rm km/h}]$, which is faster than the speed limit $V_{\rm max}^f$ of the follower. The trajectory of the leader can be seen in Fig. \ref{fig.LeaderSimulation}(a)-(b). The results of the first run are given in Fig. \ref{fig.LeaderSimulation}(c)-(d), and of the second in Fig. \ref{fig.LeaderSimulation}(e)-(f). It can be seen that in both runs, the headway between the vehicles is at least $2[{\rm s}]$ and the velocity of the follower is positive and does not exceed $100[{\rm km/h}]$, as prescribed by the guarantees.

\vspace{-10pt}
In the first run, the controller $\Sigma^{(1)}_{\rm ctr}$ chooses the maximal possible actuation input $u(k)$ guaranteeing safe behaviour given the contracts on the physical subsystem, encouraging the follower to drive as fast as possible while guaranteeing safety. For this reason, the first 100 seconds are characterised by the headway approaching $2_{\rm s}$, as the speed of the leader ($95[{\rm km/h}]$) is smaller than the speed limit for the follower. The headway grows at the last $100$ seconds of the simulation as the leader accelerates to about $105[{\rm km/h}]$, which is faster than the speed limit of the follower.
In the second simulation run, the headway grows large both in the first and the last 100 seconds, as the MPC controller $\Sigma^{(2)}_{\rm ctr}$ attempts to keep the speed of the follower around $V_{\rm des} = 90[{\rm km/h}]$.

\vspace{-7pt}
\subsection{Discussion}
\vspace{-10pt}
The numerical case study presents some of the advantages of contract theory for design in general and of the presented LP-based framework for verifying vertical contracts in particular. First, Table \ref{table.Runtime} shows the approach is scalable even for an interconnection of many components. Indeed, we verify that a collection of local contracts refines a specification on the integrated system for a network of $98$ components in about $30$ seconds, and do the same for a network of $198$ components in less than $5$ minutes. We also note that contract theory supports hierarchical design, meaning that we do not need to consider hundreds of components or subsystems at the same time. In the numerical example, it is intuitive to first consider each follower on its own, and then decompose each of them further, individually and independently from the other followers. The analysis could be carried out similarly and will have similar results. This hierarchical approach also allows different abstraction levels for each step in the hierarchy. Indeed, when defining the contract for each individual follower, we only need the variables $p_r,v_r$. The $u$-variables are only needed when bisecting each follower to its two corresponding subsystems. Moreover, variables corresponding to the measurements taken by the sensors only appear if we decompose the control subsystem into smaller components, responsible for sensing and regulation. We chose not to apply the hierarchical approach in the numerical case study but instead portray the scalability of the proposed framework. 

\vspace{-7pt}
If the networked system is designed according to the principles of contract theory, modularity is achieved by design, meaning that different components or subsystems can be analysed, designed, verified, tested, updated and replaced independently of one another. In this example, if we decide to replace a follower's controller by another control law, only the control subsystem of said follower would have to be re-verified, rather than the entire autonomous vehicle or the entire platoon. In contrast, existing formal methods that do not rely on contract theory mostly consider the entire system as one entity. Thus, any change in any component of the system must be followed by a complete re-verification process of the entire system, no matter how small the component or how insignificant the change is. In general, lack of modularity is a problem which is widespread throughout control theory, with the exception of specialised techniques like retrofit control [\cite{Ishizaki2018,Ishizaki2019,Sadamoto2017}]. As highlighted by the example, contracts allow us to prove safety of the closed-loop system before we even know the structure of each block: the same proof of safety for a piecewise-linear controller also held for an MPC-like controller.

\vspace{-10pt}
\section{Conclusion}
\vspace{-9pt}
We considered the problem of contract-based modular design for dynamical control systems. First, we extended the existing definition of contracts to incorporate situations in which the assumption on the input at time $k$ depends on the outputs up to time $k-1$, which are essential for interconnected networks with feedback. We defined contract composition for such general network interconnections, and proved the definition supports independent design of the components. We then considered vertical contracts, which are statements about the refinement of a contract on a composite system by a collection of component-level contracts. For the case of contracts defined by time-invariant inequalities, we presented efficient LP-based algorithms for verifying these vertical contracts, which scale linearly with the number of components. These results were first achieved for feedback-less networks using directed acyclic graphs, and later extended to networks with feedback interconnections but no algebraic loops using causality and strict causality. 
One possible avenue for future research is extending the presented contract-based framework to specifications defined using more general temporal logic formulae. Another direction to tackle is finding the optimal vertical contract, i.e., one is given a contract on a composite system, and the goal is to find a vertical contract which is cheapest to implement.


\vspace{-10pt}
\appendix 
\section{Proof of Theorem \ref{thm.VertCascadeN}} \label{appendix.Vert}
\vspace{-7pt}
This appendix is dedicated to proving Theorem \ref{thm.VertCascadeN}:

\vspace{-17pt}
\begin{pf}
We show that under the extendability assumptions of the theorem, the set of implications i) for all $i\in \V$ is equivalent to $\D_{\rm tot} \subseteq \D_\otimes$, and implication ii) is equivalent to $\OO_\otimes \cap \D_{\rm tot} \subseteq \OO_{\rm tot} \cap \D_{\rm tot}$. We start with the former equivalence.

\vspace{-7pt}
Suppose first that the implication i) holds for $i\in \V$, and take $(d^{\rm ext}(\cdot),y^{\rm ext} (\cdot))\in \D_{\rm tot}$. We show that $(d^{\rm ext}(\cdot),y^{\rm ext} (\cdot))\in \D_\otimes$. In other words, we show that for any $i\in \V$ and for any $\{d_j,y_j\}_{j\in \BR_+(i)}$ satisfying \eqref{eq.diDefinition} and \eqref{eq.OutputConsistency}, if $(d_j,y_j)\in \OO_j$ holds for $j\in \BR(i)$ then $(d_i,y_i)\in \OO_i$. Taking arbitrary $\{d_j,y_j\}_{j\in \BR_+(i)}$ satisfying these constraints, both $\alpha_{\rm tot}\left(\begin{smallmatrix}d^{\rm ext}(k-m_{\rm tot}^A:k) \\ y^{\rm ext} (k-m^A_{\rm tot}:k-1)\end{smallmatrix}\right) \le 0$ and $\gamma_j\left(\begin{smallmatrix}d_j(k-m_j^G:k) \\ y_j(k-m_j^G:k)\end{smallmatrix}\right) \le 0$ hold for any $k$. Thus, by applying i) for $d^{\rm ext},y^{\rm ext} ,d_j,y_j$ at times $k-m_i^A,\ldots,k$, we yield $\alpha_i\left(\begin{smallmatrix}d_i(k-m_i^A:k) \\ y_i(k-m_i^A: k-1)\end{smallmatrix}\right) \le 0$ for $k\ge m_i^A$. In particular, we have $(d_i,y_i)\in \D_i$, as claimed. As the choice of $i\in \V$ was arbitrary, we conclude that $(d^{\rm ext},y^{\rm ext} ) \in \D_{\otimes}$.

Conversely, we assume $\D_\otimes \supseteq \D_{\rm tot}$ and show the implication i) holds for $i\in \V$. We take $\{d_j,y_j\}_{j\in \BR_+(i)}, d^{\rm ext},y^{\rm ext} $ defined up to time $m_i^A$, and assume they satisfy the consistency constraints \eqref{eq.diDefinition} and \eqref{eq.OutputConsistency}, as well as

\vspace{-22pt}
\small
\begin{align*}
        &\alpha_{\rm tot}\left(\begin{smallmatrix}d^{\rm ext}(\ell-m_{\rm tot}^A:\ell) \\ y^{\rm ext} (\ell-m_{\rm tot}^A: \ell - 1)\end{smallmatrix}\right) \le 0,&& \forall \ell \in \mathcal{I}_{m_{\rm tot}^A,m_i^A},\\
        &\gamma_j\left(\begin{smallmatrix}d_j(\ell-m_j^G:\ell) \\ y_j(\ell-m_j^G:\ell)\end{smallmatrix}\right) \le 0, &&\forall \ell \in \mathcal{I}_{m_j^G,m_i^A},~j\in \BR(i).
\end{align*}\normalsize
\vspace{-20pt}

By extendibility, we find signals $\{\hat y_j,\hat d_j\}$, $\hat d^{\rm ext}$ and $\hat y^{\rm ext} $ with $\hat d^{\rm ext}(0:m_i^A) = d^{\rm ext}(0:m_i^A)$, $\hat y^{\rm ext} (0:m_i^A) = y^{\rm ext} (0:m_i^A)$, and $\hat y_j(0:m_i^A) = y_j(0:m_i^A), \hat d_j(0:m_i^A) = d_j(0:m_i^A)$ for any $j\in \BR_+(i)$, satisfying \eqref{eq.diDefinition}, \eqref{eq.OutputConsistency}, and

\vspace{-22pt}
\small
\begin{align*}
        &\alpha_{\rm tot}\left(\begin{smallmatrix}\hat d^{\rm ext}(k-m_{\rm tot}^A:k) \\ \hat y^{\rm ext} (k-m_{\rm tot}^A: k-1)\end{smallmatrix}\right) \le 0,&&\forall k\ge m_{\rm tot}^A\\
        &\gamma_j\left(\begin{smallmatrix}\hat d_j(k-m_j^G:k) \\ y_j(k-m_j^G:k)\end{smallmatrix}\right) \le 0,&&\forall k\ge m_j^G,~\forall j\in \BR(i).
\end{align*}\normalsize
\vspace{-17pt}

As $(\hat d^{\rm ext},\hat y^{\rm ext} ) \in \D_\otimes$, we conclude by Definition \ref{defn.Feedbackless} that $(d_i,y_i) \in \D_i$, i.e., that $\alpha_i\left(\begin{smallmatrix}\hat d_i(k-m_i^A:k) \\ \hat y^{\rm ext} (k-m_i^A:k-1)\end{smallmatrix}\right) \le 0$ holds for any time $k \ge m_i^A$. Taking $k=m_i^A$ gives the desired result.
We now move to the second part of the theorem, showing that the implication ii) is equivalent to  $\OO_\otimes \cap \D_{\rm tot} \subseteq \OO_{\rm tot} \cap \D_{\rm tot}$. Assume first that ii) holds, and take any $(d^{\rm ext}(\cdot),y^{\rm ext} (\cdot) \in \D_{\rm tot} \cap \OO_{\rm tot}$. By Definition \ref{defn.Feedbackless}, there exist signals $\{d_i,y_i\}_{i\in \V}$ satisfying the consistency constraints \eqref{eq.diDefinition} and \eqref{eq.OutputConsistency} and $(d_i,y_i) \in \OO_i$ for $i\in \V$. Thus, for any $k$ and $i\in \V$, both $\alpha_{\rm tot}\left(\begin{smallmatrix} d^{\rm ext}(k-m^A_{\rm tot}:k) \\ y^{\rm ext} (k-m_{\rm tot}^A: k-1)\end{smallmatrix}\right) \le 0$ and $\gamma_i\left(\begin{smallmatrix} d_i(k-m_i^G:k) \\ y_i(k-m_i^G:k)\end{smallmatrix}\right) \le 0$ hold.
The implication ii), applied to $d^{\rm ext},y^{\rm ext} ,d_i,y_i$ at times $k-m^G_{\rm tot},\ldots,k$, gives $\gamma_{\rm tot}\left(\begin{smallmatrix} d^{\rm ext}(k-m^G_{\rm tot}:k) \\ y^{\rm ext} (k-m^G_{\rm tot}:k)\end{smallmatrix}\right) \le 0$ holds for $k\ge m^G_{\rm tot}$. We thus yield $(d^{\rm ext}(\cdot),y^{\rm ext} (\cdot)) \in \OO_{\rm tot}$, as desired.

Conversely, we assume $\OO_\otimes \cap \D_{\rm tot} \subseteq \OO_{\rm tot} \cap \D_{\rm tot}$ and prove the implication ii) holds. Take $d^{\rm ext} ,y^{\rm ext} ,d_i,y_i$ defined up to time $m_{\rm tot}^G$, satisfying constraints \eqref{eq.diDefinition}, \eqref{eq.OutputConsistency}, and

\vspace{-22pt}
\small
\begin{align*}
    &\alpha_{\rm tot}\left(\begin{smallmatrix}d^{\rm ext}(0:m^A_{\rm tot}) \\ y^{\rm ext} (0:m^A_{\rm tot}-1)\end{smallmatrix}\right) \le 0,&&\forall \ell \in \mathcal{I}_{m_{\rm tot}^A,m_{\rm tot}^G}\\ &\gamma_i\left(\begin{smallmatrix}d_i(\ell-m^G_i:\ell) \\ y_i(\ell-m^G_i:\ell)\end{smallmatrix}\right) \le 0, &&\forall\ell\in \mathcal{I}_{m^G_i,m^G_{\rm tot}},~\forall i\in \V.
\end{align*}\normalsize
\vspace{-22pt}

By extendibility, we find signals $\{\hat y_i(\cdot),\hat d_i(\cdot)\}_{i\in \V}$ , $d^{\rm ext}(\cdot)$ and $y^{\rm ext} (\cdot)$, such that $\hat d^{\rm ext}(0:m_{\rm tot}) = d^{\rm ext}(0:m_{\rm tot})$, $\hat y^{\rm ext} (0:m_{\rm tot}) = y^{\rm ext} (0:m_{\rm tot})$, and both $\hat y_i(0:m_{\rm tot}) = y_i(0:m_{\rm tot}), \hat d_i(0:m_{\rm tot}) = d_i(0:m_{\rm tot})$ hold for any $i\in \V$. Moreover, for any time $k$, both the input- and output-consistency constraints \eqref{eq.diDefinition} and \eqref{eq.OutputConsistency} hold, and,

\vspace{-22pt}
\small
\begin{align*}
        &\alpha_{\rm tot}\left(\begin{smallmatrix} \hat d^{\rm ext}(k-m^A_{\rm tot}:k) \\ \hat y^{\rm ext} (k-m^A_{\rm tot}: k-1)\end{smallmatrix}\right) \le 0,&&\forall k\ge m^A_{\rm tot}~~\\
        &\gamma_i\left(\begin{smallmatrix} \hat d_i(k-m_i^G:k) \\ \hat y_i(k-m_i^G:k)\end{smallmatrix}\right) \le 0,&&\forall i\in \V, \forall k\ge m_i^G.
\end{align*}\normalsize
\vspace{-22pt}

In other words, we have $(\hat d^{\rm ext},\hat y^{\rm ext} ) \in \D_{\rm tot}$ and $(\hat d_i,\hat y_i) \in \OO_i$ for $i\in \V$. Thus, $(\hat d^{\rm ext},\hat y^{\rm ext} ) \in \D_{\rm tot} \cap \OO_\otimes \subset \D_{\rm tot} \cap \OO_{\rm tot}$, implying that $\gamma_{\rm tot}\left(\begin{smallmatrix}\hat d^{\rm ext}(k-m^G_{\rm tot}:k) \\ \hat y^{\rm ext} (k-m^G_{\rm tot}:k)\end{smallmatrix}\right) \le 0$ holds for any $k \ge m_{\rm tot}^G$. Choosing $k = m_{\rm tot}^G$ completes the proof. \qed
\end{pf}

\vspace{-15pt}
\bibliographystyle{apalike}
\bibliography{main}

\begin{thebibliography}{}

\bibitem[Baier and Katoen, 2008]{Baier2008}
Baier, C. and Katoen, J.-P. (2008).
\newblock {\em Principles of Model Checking}.
\newblock MIT press.

\bibitem[Baldwin and Clark, 2006]{Baldwin2006}
Baldwin, C.~Y. and Clark, K.~B. (2006).
\newblock Modularity in the design of complex engineering systems.
\newblock In {\em Complex engineered systems}, pages 175--205. Springer.

\bibitem[Belta et~al., 2017]{Belta2007}
Belta, C., Yordanov, B., and Gol, E.~A. (2017).
\newblock {\em Formal Methods for Discrete-Time Dynamical Systems}, volume~89.
\newblock Springer.

\bibitem[Benveniste et~al., 2018]{Benveniste2018}
Benveniste, A., Caillaud, B., Nickovic, D., Passerone, R., Raclet, J.-B.,
  Reinkemeier, P., et~al. (2018).
\newblock Contracts for system design.
\newblock {\em Foundations and Trends in Electronic Design Automation},
  12(2-3):124--400.

\bibitem[{Besselink} et~al., 2019]{Besselink2019}
{Besselink}, B., {Johansson}, K.~H., and {Van Der Schaft}, A. (2019).
\newblock Contracts as specifications for dynamical systems in driving variable
  form.
\newblock In {\em Proc. Eur. Control Conf.}, pages 263--268.

\bibitem[Blanchini and Miani, 2008]{Blanchini2008}
Blanchini, F. and Miani, S. (2008).
\newblock {\em Set-Theoretic Methods in Control}.
\newblock Springer.

\bibitem[{Chen} et~al., 2018]{Chen2018}
{Chen}, M., {Herbert}, S.~L., {Vashishtha}, M.~S., {Bansal}, S., and {Tomlin},
  C.~J. (2018).
\newblock Decomposition of reachable sets and tubes for a class of nonlinear
  systems.
\newblock {\em IEEE Trans. Autom. Control}, 63(11):3675--3688.

\bibitem[Cormen et~al., 2009]{Cormen2009}
Cormen, T.~H., Leiserson, C.~E., Rivest, R.~L., and Stein, C. (2009).
\newblock {\em Introduction to Algorithms}.
\newblock MIT press.

\bibitem[Desoer and Vidyasagar, 2009]{Desoer2009}
Desoer, C.~A. and Vidyasagar, M. (2009).
\newblock {\em Feedback Systems: Input-Output Properties}.
\newblock SIAM.

\bibitem[{Eqtami} and {Girard}, 2019]{Eqtami2019}
{Eqtami}, A. and {Girard}, A. (2019).
\newblock A quantitative approach on assume-guarantee contracts for safety of
  interconnected systems.
\newblock In {\em Proc. Eur. Control Conf.}, pages 536--541.

\bibitem[Ghasemi et~al., 2020]{Ghasemi2020}
Ghasemi, K., Sadraddini, S., and Belta, C. (2020).
\newblock Compositional synthesis via a convex parameterization of
  assume-guarantee contracts.
\newblock In {\em Proc. 23rd Int. Conf. Hybrid Syst.: Comput. Control}, pages
  1--10.

\bibitem[Huang and Kusiak, 1998]{Huang1998}
Huang, C.-C. and Kusiak, A. (1998).
\newblock Modularity in design of products and systems.
\newblock {\em IEEE Trans. Syst., Man, Cybern.}, 28(1):66--77.

\bibitem[Ishizaki et~al., 2019]{Ishizaki2019}
Ishizaki, T., Kawaguchi, T., Sasahara, H., and Imura, J.-i. (2019).
\newblock Retrofit control with approximate environment modeling.
\newblock {\em Automatica}, 107:442--453.

\bibitem[Ishizaki et~al., 2018]{Ishizaki2018}
Ishizaki, T., Sadamoto, T., Imura, J.-i., Sandberg, H., and Johansson, K.~H.
  (2018).
\newblock Retrofit control: Localization of controller design and
  implementation.
\newblock {\em Automatica}, 95:336--346.

\bibitem[{Meyer}, 1992]{Meyer1992}
{Meyer}, B. (1992).
\newblock Applying 'design by contract'.
\newblock {\em Computer}, 25(10):40--51.

\bibitem[{Nuzzo} et~al., 2015]{Nuzzo2015}
{Nuzzo}, P., {Sangiovanni-Vincentelli}, A.~L., {Bresolin}, D., {Geretti}, L.,
  and {Villa}, T. (2015).
\newblock A platform-based design methodology with contracts and related tools
  for the design of cyber-physical systems.
\newblock {\em Proc. IEEE}, 103(11):2104--2132.

\bibitem[{Nuzzo} et~al., 2014]{Nuzzo2014}
{Nuzzo}, P., {Xu}, H., {Ozay}, N., {Finn}, J.~B., {Sangiovanni-Vincentelli},
  A.~L., {Murray}, R.~M., {Donzé}, A., and {Seshia}, S.~A. (2014).
\newblock A contract-based methodology for aircraft electric power system
  design.
\newblock {\em IEEE Access}, 2:1--25.

\bibitem[Rantzer, 2015]{Rantzer2015}
Rantzer, A. (2015).
\newblock Scalable control of positive systems.
\newblock {\em Eur. J. Control}, 24:72--80.

\bibitem[Sadamoto et~al., 2017]{Sadamoto2017}
Sadamoto, T., Chakrabortty, A., Ishizaki, T., and Imura, J.-i. (2017).
\newblock Retrofit control of wind-integrated power systems.
\newblock {\em IEEE Trans. Power Syst.}, 33(3):2804--2815.

\bibitem[Saoud et~al., 2018a]{Saoud2018}
Saoud, A., Girard, A., and Fribourg, L. (2018a).
\newblock On the composition of discrete and continuous-time assume-guarantee
  contracts for invariance.
\newblock In {\em Proc. Eur. Control Conf.}, pages 435--440.

\bibitem[Saoud et~al., 2021]{Saoud2021}
Saoud, A., Girard, A., and Fribourg, L. (2021).
\newblock Assume-guarantee contracts for continuous-time systems.
\newblock {\em Automatica}, 134:109910.

\bibitem[Saoud et~al., 2018b]{Saoud2018b}
Saoud, A., Jagtap, P., Zamani, M., and Girard, A. (2018b).
\newblock Compositional abstraction-based synthesis for cascade discrete-time
  control systems.
\newblock In {\em Proc. 6th IFAC Conf. Anal. Des. Hybrid Syst.}, pages 13--18.

\bibitem[Shali et~al., 2021]{Shali2021}
Shali, B., van~der Schaft, A., and Besselink, B. (2021).
\newblock Behavioural contracts for linear dynamical systems: Input assumptions
  and output guarantees.
\newblock In {\em Proc. Eur. Control Conf.}, pages 564--569.

\bibitem[Sharf et~al., 2021a]{Sharf2021b}
Sharf, M., Besselink, B., and Johansson, K.~H. (2021a).
\newblock Verifying contracts for perturbed control systems using linear
  programming.
\newblock {\em arXiv preprint arXiv:2111.01259}.

\bibitem[Sharf et~al., 2021b]{SharfADHS2020}
Sharf, M., Besselink, B., Molin, A., Zhao, Q., and Johansson, K.~H. (2021b).
\newblock {Assume/Guarantee} contracts for dynamical systems: Theory and
  computational tools.
\newblock In {\em Proc. 7th IFAC Conf. Anal. Des. Hybrid Syst.}

\bibitem[{\v{S}}iljak and Ze{\v{c}}evi{\'c}, 2005]{Siljak2005}
{\v{S}}iljak, D.~D. and Ze{\v{c}}evi{\'c}, A. (2005).
\newblock Control of large-scale systems: Beyond decentralized feedback.
\newblock {\em Annu. Rev. Control}, 29(2):169--179.

\bibitem[{Smith} et~al., 2016]{Smith2016}
{Smith}, S.~W., {Nilsson}, P., and {Ozay}, N. (2016).
\newblock Interdependence quantification for compositional control synthesis
  with an application in vehicle safety systems.
\newblock In {\em Proc. IEEE Conf. Decision Control}, pages 5700--5707.

\bibitem[Tabuada, 2009]{Tabuada2009}
Tabuada, P. (2009).
\newblock {\em Verification and Control of Hybrid Systems: a Symbolic
  Approach}.
\newblock Springer Science \& Business Media.

\bibitem[Ulrich, 1995]{Ulrich1995}
Ulrich, K. (1995).
\newblock The role of product architecture in the manufacturing firm.
\newblock {\em Research policy}, 24(3):419--440.

\bibitem[Willems, 1972a]{Willems1972a}
Willems, J.~C. (1972a).
\newblock Dissipative dynamical systems part i: General theory.
\newblock {\em Archive for Rational Mechanics and Analysis}, 45(5):321--351.

\bibitem[Willems, 1972b]{Willems1972b}
Willems, J.~C. (1972b).
\newblock Dissipative dynamical systems part ii: Linear systems with quadratic
  supply rates.
\newblock {\em Archive for Rational Mechanics and Analysis}, 45(5):352--393.

\bibitem[{Zamani} and {Arcak}, 2018]{Zamani2018}
{Zamani}, M. and {Arcak}, M. (2018).
\newblock Compositional abstraction for networks of control systems: A
  dissipativity approach.
\newblock {\em IEEE Trans. Control Netw. Syst.}, 5(3):1003--1015.

\end{thebibliography}

\end{document}